\documentclass[aps,prb,epsfig,preprint,superscriptaddress,showpacs]{revtex4-2}
\usepackage{epstopdf}
\usepackage{graphicx}  
\usepackage{float}
\usepackage{dcolumn}   
\usepackage{bm}        
\usepackage{amssymb}
\usepackage[english]{babel}
\usepackage[english]{babel}

\begin{document}

\title{Observation of robust spin-phonon coupling and indication of hidden structural transition in the spin-driven ferroelectrics Mn$_4${\bf \textit{B}}$_2$O$_9$ ({\bf \textit{B}} = Nb, Ta)}

\author{Rajesh Jana}
\email[Contact author: ]{rajesh.jana@hpstar.ac.cn}
\affiliation{Center for High Pressure Science and Technology Advanced Research (HPSTAR), Beijing 100193, P. R. China}
\affiliation{Solid State Physics Division, Bhabha Atomic Research Centre, Mumbai 400085, India}

\author{Alka Garg}
\affiliation{High Pressure and Synchrotron Radiation Physics Division, Bhabha Atomic Research Centre, Mumbai 400085, India}
\affiliation{Homi Bhabha National Institute, Anushaktinagar, Mumbai 400094, India}
\author{Mayuresh Mukadam}
\affiliation{Solid State Physics Division, Bhabha Atomic Research Centre, Mumbai 400085, India}

\author{Rekha Rao}
\email[Contact author: ]{rekhar@barc.gov.in}
\affiliation{Solid State Physics Division, Bhabha Atomic Research Centre, Mumbai 400085, India}
\affiliation{Homi Bhabha National Institute, Anushaktinagar, Mumbai 400094, India}

\author{Thomas Meier}
\email[Contact author: ]{thomasmeier@sharps.ac.cn}
\affiliation{Shanghai Key Laboratory MFree, Institute for Shanghai Advanced Research in Physical Sciences, Shanghai 201203, P. R. China}
\affiliation{Center for High Pressure Science and Technology Advanced Research (HPSTAR), Beijing 100193, P. R. China}

\begin{abstract}

We report detailed Raman spectroscopic and magnetic susceptibility studies on the spin-driven ferroelectric compounds Mn$_4$Nb$_2$O$_9$ (MNO) and Mn$_4$Ta$_2$O$_9$ (MTO). Both systems exhibit strong spin–phonon coupling below the short-range magnetic ordering temperature ($T_{sro}\sim223$ K), followed by further renormalization of several Raman modes at the long-range magnetic ordering temperatures ($T_N$ = 120 K for MNO and 110 K for MTO). Pronounced anomalies in Raman mode frequencies and linewidths, along with the emergence of octahedral modes between $T_{sro}$ and $T_N$, indicate a possible low-symmetry structural transition, more evident in MNO and closely linked to magnetic ordering in MTO. Distinct low-temperature evolutions of Raman mode shift, linewidth, and integrated intensity in MNO and MTO highlight the role of the nonmagnetic $B$-site cation in tuning spin–lattice coupling, driven by differences in spin–orbit coupling and orbital hybridization between Nb$^{5+}$ (4$d$) and Ta$^{5+}$ (5$d$). By combining Raman spectroscopy with nuclear magnetic resonance, and diffuse reflectance spectroscopy, we further show that Mn-based systems possess a more distorted local structure than their Co analogues, while their electronic structures differ despite comparable band gaps. These results provide a comprehensive understanding of spin–lattice coupling in Mn- and Co-based $A_4B_2$O$_9$ magnetoelectric systems.

\end{abstract}

\maketitle
\section{Introduction}

	Spin-driven ferroelectric materials, such as type-II multiferroics (MFs) and linear magnetoelectric (LME) systems, have attracted considerable interest owing to their ability to achieve efficient mutual control between magnetic and electric order parameters within the magnetically ordered phase—an essential feature for next-generation multifunctional technologies. In type-II MFs, ferroelectricity and magnetoelectric (ME) coupling emerge spontaneously from specific magnetic structures that break spatial inversion symmetry. In contrast, LME systems do not possess spontaneous ferroelectricity in their ground state; instead, electric polarization is induced only upon the application of an external magnetic field. However, the strong magnetic frustration typically present in type-II MFs tends to suppress their magnetic ordering temperatures and hence ME properties, thereby constraining their practical applicability. LME systems, on the other hand, generally experience weaker magnetic frustration, enabling the persistence of ME coupling at comparatively higher temperatures.  

Although the family of LME materials $A$$_4$$B$$_2$O$_9$ ($A$ = Fe, Co, Mn; $B$ = Nb, Ta) was first synthesized and studied in 1961 \cite{Bertaut}, it has recently attracted renewed interest due to the discovery of spin-driven and magnetically controllable dielectric responses and ferroelectricity, electric-field modulation of magnetization, and spin-lattice interactions, all together demonstrating strong magnetoelectric and magnetoelastic couplings below their magnetic ordering temperatures \cite{Maignan, Panja, Mehra, Fang, Zheng, Datta, Goel}. These compounds crystallize in the trigonal \textit{P-3c1} space group and exhibit antiferromagnetic ordering characterized by antiferromagnetic interactions within the basal \text{ab}-plane \cite{Jana, Narayanan}. This arrangement gives rise to ferromagnetic chains along the \textit{c}-axis, although the spin orientations vary depending on the magnetic \textit{A}-site ion. In Mn$_4$$B$$_2$O$_9$, the magnetic moments on the two inequivalent $A$-sites are collinearly aligned along the \textit{c}-axis \cite{Narayanan, Deng}, whereas the Co- and Fe-based analogues display canted spin structures with spins predominantly confined within the basal plane. The canting angle is significantly larger in the Co compounds compared to their Fe counterparts \cite{Deng2, Choi, Ding}. Consequently, the variation in $A$-site magnetic configurations across the $A$$_4$$B$$_2$O$_9$ family gives rise to distinct microscopic origins of magnetoelectric coupling. In Co-based systems, ME coupling originates from the inverse Dzyaloshinskii–Moriya (DM) interaction (or spin-current mechanism), which induces ferroelectric polarization—consistent with their canted spin structure \cite{Deng2}. By contrast, in Mn-based systems, ME coupling originates primarily from magnetic exchange striction associated with the collinear spin arrangement, while in Fe$_4$Nb$_2$O$_9$ (FNO), it originates from magnetostriction and spin-dependent \textit{d–p} hybridization \cite{Zheng, Deng, Ding, Zhang}.
	
$B$-site variation also leads to distinct physical behaviors across this family. In Fe$_4$$B$$_2$O$_9$, the Ta analogue exhibits genuine multiferroic behavior with multiple magnetoelectric transitions, compared to only two transitions in the Nb counterpart, indicating stronger magneto-lattice coupling and highlighting the crucial role of the nonmagnetic $B$-site cation in influencing the magnetic sublattice \cite{Panja2, Maignan2}. A similar trend is observed in Mn$_4$$B$$_2$O$_9$, where the Ta analogue (MTO) displays stronger magnetic exchange striction around both short-range and long-range magnetic ordering compared to the Nb analogue (MNO) \cite{Narayanan}. In Co$_4$$B$$_2$O$_9$, our earlier works revealed enhanced spin–phonon coupling below the short-range magnetic ordering temperature and distinct pressure-induced phase transitions in the Ta analogue \cite{Jana2, Jana3}. Likewise, Park et al. reported stronger spin–phonon coupling below the long-range antiferromagnetic transition in the Ta-based compound, attributed to the enhanced spin–orbit coupling (SOC) associated with the heavier Ta cation \cite{Park}. Furthermore, superior magnetothermal conductivity in Co$_4$Ta$_2$O$_9$ (CTO) compared to Co$_4$Nb$_2$O$_9$ (CNO) has also been reported, which was ascribed to enhanced magnetic and phonon scattering interactions \cite{Ueno}.
	
Within the $A$$_4$$B$$_2$O$_9$ family, the Mn-based compounds exhibit the highest antiferromagnetic ordering temperatures, with T$_N$ $\sim$ 109 K for the Nb analogue and T$_N$ $\sim$ 102 K for the Ta analogue \cite{Zheng, Narayanan}. In these strongly coupled magnetoelectric systems, significant modulation of phonon properties is expected in response to changes in magnetic ordering via spin–phonon coupling (SPC), which plays a crucial role in spintronic, memory, and multifunctional electronic devices. The observed strong SPC in Co-based analogues, both near the long-range magnetic ordering temperature and below the temperatures associated with short-range magnetic correlations, further reinforces this expectation \cite{Park, Jana2, Jana3}.  Therefore, a clear understanding of SPC is essential in coupled ferroic materials, as the interaction between different ferroic order parameters is often mediated via phonons, providing key insights into the fundamental coupling mechanisms \cite{Tang}. A previous low-temperature Raman study on single-crystalline MNO investigated the relative changes in all Raman mode frequencies in the 83–283 K range to probe possible spin–phonon coupling around T$_N$ \cite{Chen}. However, examining all modes collectively limited the ability to resolve subtle, mode-dependent anomalies, and no clear phonon-frequency renormalization was reported. Furthermore, that work did not investigate mode linewidths or spectral weights (integrated intensities), which are essential critical indicators for probing SPC \cite{Son, Roy, Cottam, Rao}.	
	
Nevertheless, evidence of a strong magnetostrictive effect has been reported in both Mn$_4$Nb$_2$O$_9$ (MNO) and Mn$_4$Ta$_2$O$_9$ (MTO) around magnetic ordering temperatures, suggesting the existence of spin–phonon coupling (SPC) in these systems \cite{Cao, Narayanan, Panja}. This highlights the need for a more detailed mode-dependent and systematic temperature-dependent vibrational study to examine these materials. Notably, no vibrational study on MTO has yet been reported to explore the spin–phonon interaction at low temperatures, despite the expectation that orbital hybridization between the spatially extended Ta 5$d$ and Mn 3$d$ orbitals as well as the O 2$p$ states would substantially influence the structural, magnetic, and vibrational properties of MTO compared to its Nb analogue, which involves the relatively less extended Nb 4$d$ orbitals \cite{Ueno, Solovyev}. A thorough understanding of SPC in such systems is not only crucial for unraveling the fundamental interplay among spin, lattice, and orbital degrees of freedom but also for enabling the stabilization of desired magnetic states through external perturbations such as epitaxial strain, pressure, magnetic fields, or temperature modulation \cite{Rao, Rodríguez-Hernández, Carlisle, Pawbake}.

Furthermore, a proper understanding of structural symmetry breaking—either at the local or average scale—is vital for achieving efficient magnetoelectric coupling. Within the $A$$_4$$B$$_2$O$_9$ family, such structural symmetry lowering has so far been unambiguously reported only in FNO, which undergoes a transition from a trigonal \textit{P-3c1} to a monoclinic \textit{C2/c} phase just below $T_N$, as revealed by neutron and x-ray diffraction studies \cite{Jana}. However, conventional diffraction techniques may overlook subtle or local structural distortions; thus, the use of local probes such as Raman spectroscopy becomes essential to uncover possible low-temperature structural modifications in MNO and MTO.   

In this work,  we have employed the capabilities of low-laser-power, temperature-dependent Raman spectroscopy, which serves as a highly sensitive probe to study the influence of magnetic interactions on phonon properties; and structural symmetry. Our detailed analysis of individual Raman modes for both systems, complemented by magnetic susceptibility measurements, reveals pronounced renormalization of phonon frequency, linewidth, and integrated intensity around the magnetic ordering temperatures. Signatures of local and average symmetry changes are also evident from the emergence of new Raman modes. Additionally, comparative studies using Raman spectroscopy, diffuse reflectance spectroscopy, and nuclear magnetic resonance  provide comprehensive insights into the structural, vibrational, and electronic properties of Co- and Mn-based $A$$_4$$B$$_2$O$_9$ systems, as well as their mutual correlations. 

\section{Experimental details}

\subsection{Sample synthesis and phase characterization}
	  
High-quality polycrystalline samples of MNO and MTO are synthesized via the conventional solid-state reaction route. Stoichiometric amounts of Mn$_2$O$_3$, Nb$_2$O$_5$ and Ta$_2$O$_5$ are preheated at 300$^{\circ}$C for 3 h to remove moisture, thoroughly mixed in an agate mortar and pestle for several hours and then pressed into discs. These discs are calcined at 1000$^{\circ}$C for 24 h in an argon atmosphere. The resulting products are reground and re-presed and subsequently calcinated at 1200$^{\circ}$C  multiple times with intermediate grinding to suppress secondary phase formation and enhance sample homogeneity. Structural characterization at ambient conditions is conducted using a rotating-anode X-ray diffractometer with Cu K$_\alpha$ radiation. The synthesis of CNO and CTO in similar method is reported in our previous studies \cite{Jana2, Jana3}.
		
\subsection{Diffuse reflectance spectroscopy}		
	
Diffuse reflectance spectra (DRS) for MNO and MTO, along with their Co analogues, are recorded at room temperature using a fibre optic based UV-Vis-NIR spectrometer equipped with a CCD (charge coupled device) detector. Balanced deuterium tungsten source covering a broad range of electromagnetic spectrum from 210-2500 nm is used for sample illumination. Spectralon, a sintered PTFE-based material from Labsphere, is used as a reference material to correct for baseline shifts as it is the whitest substance, reflecting more than 99\% of the light. The recorded diffused reflectance data are converted to absorbance using the Kubelka–Munk function for subsequent optical band gap analysis.
			
\subsection{Nuclear magnetic resonance spectroscopy}
$^{93}$Nb nuclear magnetic resonance (NMR) measurements are performed on MNO and CNO at room temperature under a magnetic field of 9.3 T. An in-house-built NMR probe with a customized \textit{L–C} resonator is used to tune the radio-frequency excitation to the Nb Larmor frequency of 10.42 MHz/T. Frequency-sweep NMR spectra are obtained by exciting narrow frequency windows of 0.4-0.8 MHz and repeating the measurements over a range of 92–102 MHz range, as the spectra for both samples span approximately 10 MHz. The spin–lattice relaxation time (T1) is determined using a saturation recovery with solid echo read-out, while the spin–spin relaxation time (T2) is extracted from measurements of the longitudinal magnetization decay at different pulse delay intervals.

\subsection{Low-temperature Raman spectroscopy}		
Raman spectra at ambient and low temperatures are collected in backscattering geometry using a Horiba Yvon LabRam HR Evolution spectrometer with a 600 g/mm grating and a 532 nm excitation laser. The laser power is kept below 0.5 mW to minimize local heating. Low temperatures down to 77 K are achieved using a Linkam THMS600 stage with 0.1 K resolution. At each temperature, spectra are recorded after a 8 min stabilization period to ensure thermal equilibrium, while maintaining the same collection spot. A 20x long-working-distance, infinity-corrected objective is used to focus the unpolarized excitation laser and collect the scattered signal through the temperature stage’s optical window. 

\subsection{Low-temperature magnetic susceptibility measurement} 
Magnetic susceptibility is measured in zero-field-cooled (ZFC) and field-cooled (FC) modes over 2–300 K using a Vibrating Sample Magnetometer (VSM, Cryogenic Ltd., UK). In ZFC mode, samples are cooled to 2 K without a magnetic field, after which data recorded upon warming under a magnetic field of 660 Oe for MNO and 940 Oe for MTO. In FC mode, the same field is applied during cooling, and susceptibility is recorded on warming.

\section{Results and Discussion}

\subsection{Ambient characterization}

The Rietveld-refined XRD patterns of MNO and MTO at room temperature are presented in Fig. 1(a)-(b). All observed Bragg reflections for both compounds are well indexed to the trigonal \textit{P-3c1} symmetry, confirming the formation of single-phase materials. The unit cell is consist of alternating nearly planar (L1) and buckled (L2) honeycomb layers stacked along the \textit{c} axis, as illustrated in Fig. 1(c)-(e). The refined lattice parameters, along with those of the Co-based analogues CNO \cite{Jana2} and CTO \cite{Jana3} from our previous studies, are summarized in Table I. The Mn-based compounds exhibit larger lattice parameters and unit cell volumes than their Co counterparts, consistent with the larger ionic radius of Mn$^{2+}$ compared to Co$^{2+}$. A slight increase in cell volume is also observed for $B$-site variation from Nb$^{5+}$ to Ta$^{5+}$. Interestingly, despite expansion in unit cell volume of MNO and MTO upon replacing Co with Mn, the $c/a$ raio decreases, while slight increase in $c/a$ ratio is observed for the Ta analogues compared to their Nb counterparts (Table I).

\begin{figure*}[ht!]
\includegraphics[width=18cm]{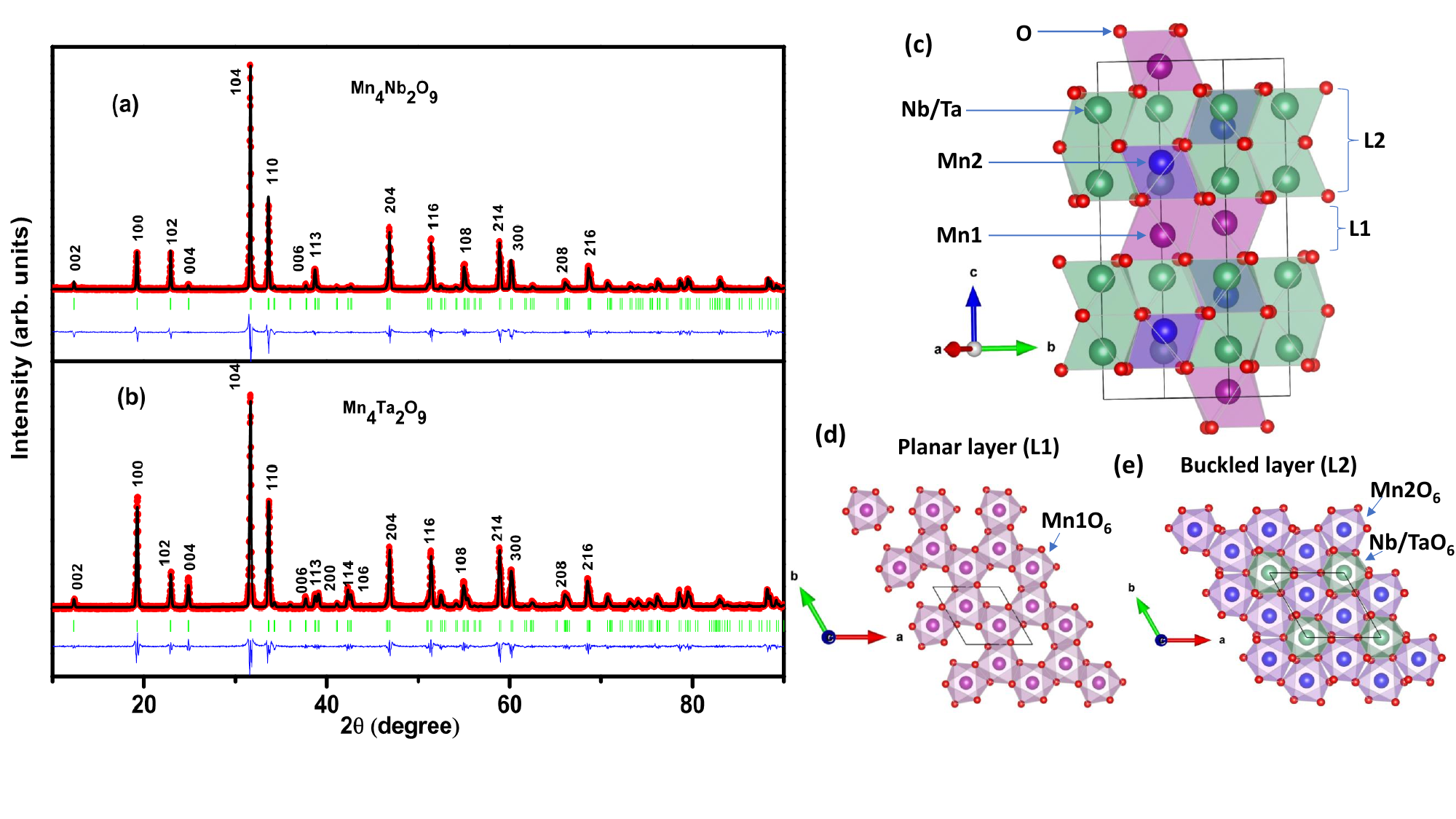}
\caption{\label{fig1} Rietveld-refined X-ray diffraction patterns of (a) MNO and (b) MTO at ambient pressure. Experimental data are shown as red circles, and the solid black lines represent the calculated profiles from the refinement. The difference between the observed and calculated patterns is displayed as blue curves at the bottom of each panel. Vertical green tick marks indicate the allowed Bragg reflection positions for the trigonal phase. (c) Schematic illustration of the trigonal \textit{P-3c1} unit cell of MNO/MTO. (d) Planar honeycomb layer (L1) formed by edge-sharing Mn1O$_6$ octahedra when viewed along the \textit{c} axis. (e) Top view of the buckled honeycomb layer (L2), highlighting the alternating connectivity of magnetic Mn2O$_6$ and nonmagnetic octahedra Nb/TaO$_6$.} 
\end{figure*}

		\begingroup
 \squeezetable	
		\begin{table*}[htb]
		 \squeezetable
\caption{\label{tab:table1
}
Comparison of lattice parameters of MNO, MTO and their Co counterparts \cite{Jana2, Jana3} crystallize in trigonal \textit{P-3c1} space group (No. 165, Z = 2)  along with mass, ionic radii \cite{Shannon}, oxidation state, spin state and coordination number of the constituent elements.}
\begin{ruledtabular}
\begin{tabular}{cccccccccccc}
 Material & \textit{a} (\AA) & \textit{c} (\AA) & \textit{c}/\textit{a} & \textit{V} (\AA$^3$)  &  element & mass (u) & ionic radii (\AA) & oxidation state & spin sate & coordination \\
\hline
 CTO  & 5.1723(6) & 14.1494(3) & 2.7356 & 327.82(5)  & Co & 58.933 & 0.745 & +2 & high & 6\\

 CNO  & 5.1693(4) & 14.1267(5) & 2.7328 & 326.91(4)  & Mn & 54.938 & 0.83 & +2 & high & 6\\

 MTO & 5.3308(2) & 14.3388(4) & 2.6898 & 352.88(4)   & Ta & 180.948 & 0.64 & +5 & - & 6  \\

 MNO & 5.3265(3) & 14.3076(6) & 2.6861 & 351.54(5)   & Nb & 92.906 & 0.64 & +5 & - & 6 \\

\end{tabular}
\end{ruledtabular}
\end{table*}
		
Ambient Raman spectra of MNO and MTO, together with their Co analogues, are presented in Fig. 2, and the corresponding Raman mode assignments following previous report \cite{Chen} are listed in Table II. As all four compounds are synthesized by the similar procedure and their Raman spectra are collected under identical conditions (instrument, grating, laser), a direct comparison of their Raman modes provides meaningful insight into their lattice vibrational properties. The factor group analysis performed for the \textit{P-3c1} structure of the $A$$_4$$B$$_2$O$_9$ compounds, comprising 30 atoms per primitive cell, predicts a total of 22 Raman-active modes (7A$_{1g}$ + 15E$_g$) \cite{Park, Jana2}. In our experiments, 17 and 16 Raman-active modes are observed for MNO and MTO, respectively, consistent with the 17 modes reported previously for MNO single crystals \cite{Chen}, with comparable frequency positions. In contrast, only 11 and 15 modes are detected for the Co-based analogues CNO and CTO, respectively, indicating a higher degree of local structural distortion in the Mn-based systems compared to their Co counterparts. Raman mode frequencies are governed by the force constants ($k$) and the reduced masses ($m$) of the constituent atoms ($\omega$ $\sim$ ($k/m$)$^{1/2}$), which are influenced by the ionic radii and atomic masses of the elements, respectively. To aid in understanding the variations in mode frequencies across the four systems, the relevant ionic radii and atomic masses are summarized in Table I.
		
		\begin{figure*}[ht!]
\includegraphics[width=18cm]{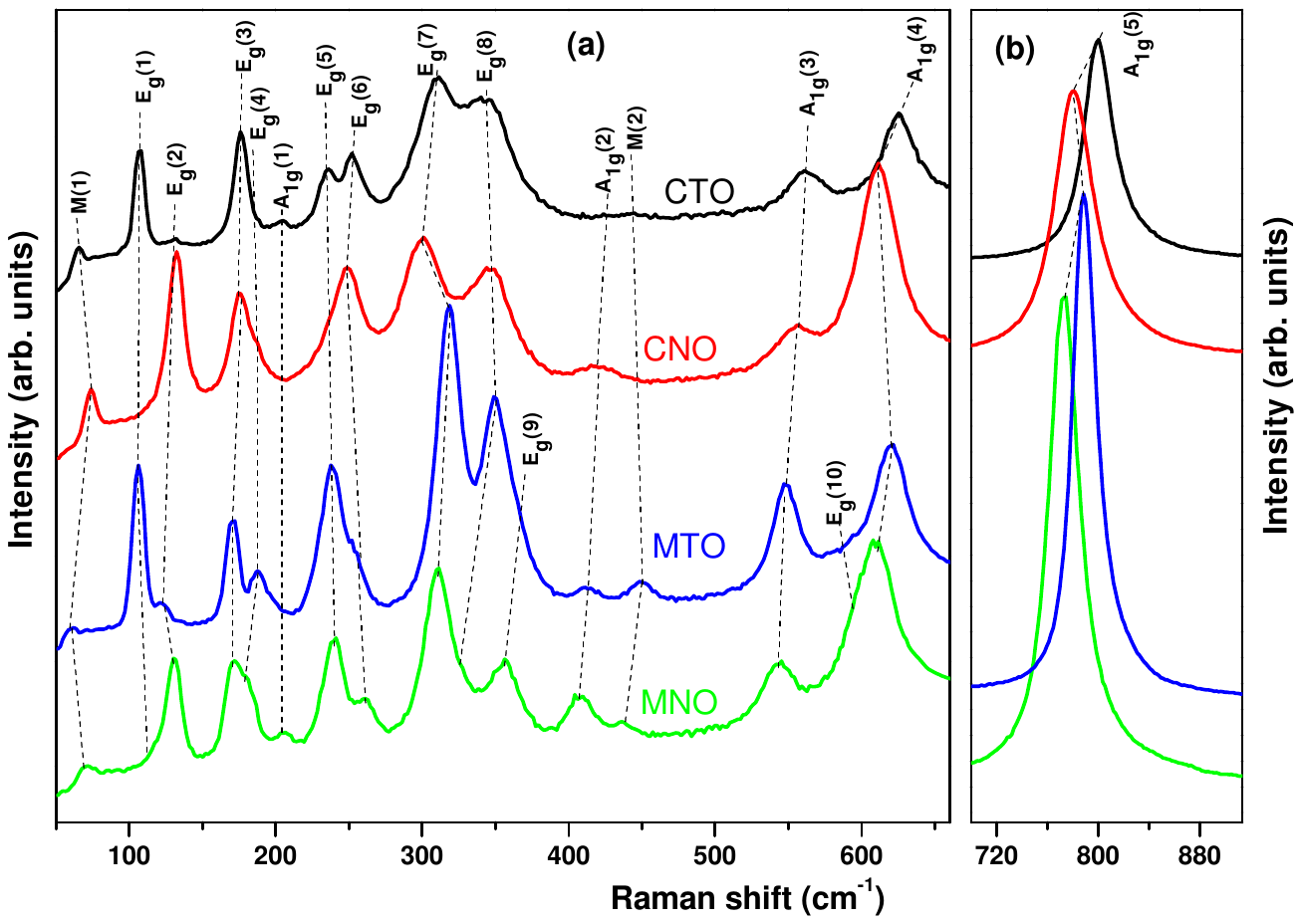}
\caption{\label{fig2} Raman spectra of MNO and MTO compared with their Co-based analogues in the ranges (a) 50–680 cm$^{-1}$ and (b) 680–900 cm$^{-1}$. Black dashed lines indicate the systematic evolution of corresponding Raman modes across the four compounds.} 
\end{figure*}

\begingroup
 \squeezetable	
		\begin{table*}[htb]
		 \squeezetable
\caption{\label{tab:table2
}
Raman mode frequencies ($cm^{-1}$) of MNO and MTO as well as their Co variants \cite{Jana2, Jana3} at ambient temperature.}
\begin{ruledtabular}
\begin{tabular}{cccccccccccccccccc}
 Material & M(1)  & E$_{g}$(1) & E$_{g}$(2) & E$_{g}$(3) & E$_{g}$(4) & A$_{1g}$(1)  & E$_{g}$(5)  & E$_{g}$(6) & E$_{g}$(7) & E$_{g}$(8) & E$_{g}$(9) & A$_{1g}$(2) & M(2) & A$_{1g}$(3) & E$_{g}$(10) & A$_{1g}$(4) & A$_{1g}$(5)\\
\hline
CTO & 65 & 106 & 132  & 174 & 189 & 205 & 234 & 250 & 310 & 345 &    & 424 & 443 & 562 &     &622 & 799 \\ 

CNO & 73 &     & 131  & 175  & 185 &       &      & 247 & 299 & 346 &   & 420 &     & 553 &     & 610 & 781 \\ 

MTO & 60 & 106 & 125  & 170 & 189 &  & 238 & 253 & 318 & 349 & 358   & 414 & 450 & 548 & 588 & 620 & 789 \\

MNO & 67 & 117 & 130  & 170  & 181 & 206 & 239 & 262 & 311 & 326 & 357 & 407 & 440 & 544 & 598 & 612 & 773 \\ 
						
\end{tabular}
\end{ruledtabular}
\end{table*}

The larger ionic radius of Mn$^{2+}$ results in an expanded unit cell for the Mn-based compounds compared to their Co-based counterparts. This expansion is expected to shift Raman modes to lower frequencies due to a reduction in the force constant. The mass effect of Mn$^{2+}$, however, is relatively minor, as its atomic mass is only slightly lower than that of Co. In contrast, replacing Nb$^{5+}$ with Ta$^{5+}$ at the $B$-site is expected to have a stronger influence on lattice vibrations owing to the substantially heavier mass of Ta. Additionally, other factors such as spin–orbit coupling (SOC) and orbital hybridization may also play important roles in modifying the Raman spectra. Considering these aspects, we analyze the variations in individual Raman modes arising from changes at both the $A$-site and $B$-site.

The lowest wavenumber mode, M(1), is influenced by both $A$-site and $B$-site atoms. In Co-based systems, this mode shifts from 73  to 65 cm$^{-1}$ when Nb is replaced by the heavier Ta. A similar trend is observed for Mn-based systems, where it shifts from 67 cm$^{-1}$ in Nb to 60 cm$^{-1}$ in Ta. This shift is primarily attributed to the mass effect of Nb and Ta. The larger unit cell of Mn-based systems also results in lower frequencies for both Mn-Nb and Mn-Ta compounds. The E$_g$(1) mode, attributed to spin-orbit coupling-induced structural distortion (as reported in Ref. \cite{Jana2}), is observed with comparable intensity in both CTO and MTO at 106 cm$^{-1}$. While its intensity becomes extremely low in MNO and completely absent in CNO. The absence or weak intensity of this mode in Nb-based systems and its intense presence in both CTO and MTO at same positions suggests that it is primarily dependent on $B$-site Ta displacement. The E$_g$(2) mode frequency remains nearly constant at around 131–132 cm$^{-1}$ in the Co analogues, whereas in the Mn-based systems it shifts toward lower frequency, from 130 cm$^{-1}$ for the Nb compound to 125 cm$^{-1}$ for the Ta variant. Interestingly, the intensity of this mode becomes significantly weaker in CTO and MTO, indicating a pronounced dependence on the \textit{B}-site cation.

The E$_g$(3) mode is detected at 175 cm$^{-1}$ for CNO, 174 cm$^{-1}$ for CTO, and 170 cm$^{-1}$ for both MNO and MTO. This suggests a predominant dependence on $A$-site atomic movement, with frequency variations related to the ionic radii of $A$-site atoms. The E$_g$(4) mode shows a dominant dependence on the $B$-site cation. A particularly interesting observation is the behavior of the A$_{1g}$(1) mode, which is present only in Nb compound when Mn is at the $A$-site but in Ta compound when Co is at the $A$-site, occurring at nearly the same position. The E$_g$(5) and E$_g$(6) modes exhibit another intriguing behavior. In CNO, the Eg$_(5)$ mode merges completely with Eg$_(6)$, whereas in CTO they are clearly resolved. Furthermore, the E$_g$(6) mode shifts to higher frequency, from 247 to 250 cm$^{-1}$, when Nb is replaced by Ta, contrary to the trend expected from the ionic radii effect. In the Mn-based systems, an entirely opposite behavior is observed. The two modes are well resolved when Nb occupies the $B$-site, but only slightly distinguishable in the Ta analogue. Nevertheless, both modes undergo a red shift in frequency upon Nb-to-Ta substitution (MNO → MTO), consistent with the expected \textit{B}-site mass effect. The contrasting evolution of mode splitting and the opposite direction of the E$_g$(6) frequency shifts between Mn- and Co-based systems upon $B$-site variation indicate a more intricate lattice vibration mechanism, likely influenced by spin–orbit coupling (SOC), orbital hybridization, and associated local structural modifications.

Most fascinating behaviour is observed in the frequency range of 270-400 cm$^{-1}$. In this region, two modes, E$_g$(7) and E$_g$(8), are detected for the Co-based systems. The E$_g$(7) mode shifts to higher frequency, from 299 cm$^{-1}$ in CNO to 310 cm$^{-1}$ in CTO, while the E$_g$(8) mode remains nearly unchanged around 345–346 cm$^{-1}$. In stark contrast, three modes, E$_g$(7)–E$_g$(9), are observed for the Mn-based systems within the same frequency range. In MNO, the E$_g$(8) mode at 326 cm$^{-1}$ appears merged with its lower-frequency neighbor E$_g$(7) (311 cm$^{-1}$), whereas E$_g$(9) (357 cm$^{-1}$) is well resolved. In MTO, however, the E$_g$(9) (358 cm$^{-1}$) mode overlaps with the E$_g$(8) (349 cm$^{-1}$), while E$_g$(7) (318 cm$^{-1}$) remains distinct. These overlapping modes exhibit clear separation at lower temperatures, as discussed later in Sections {\bf D} and {\bf E}. Overall, the E$_g$(8) and E$_g$(9) modes of MNO/MTO exhibit a pronounced transfer of intensity, while the E$_g$(7) mode of CNO/CTO and both E$_g$(7) and E$_g$(8) modes of the Mn-based systems undergo substantial blue shifts as the $B$-site is varied from Nb to Ta. Despite the nearly identical ionic radii of Nb$^{5+}$ and Ta$^{5+}$, this behavior highlights the critical role of additional factors, including orbital hybridization between Mn/Co 3$d$, $B$-site $d$ (Nb 4$d$/Ta 5$d$), and O 2$p$ orbitals, variations in local bond covalency and bond lengths, and spin–orbit coupling from the heavier Ta ion. These factors collectively modify the force constants and lattice dynamics, strongly influencing the observed vibrational behavior.

The A$_{1g}$(2) mode exhibits a red shift upon substituting Co with Mn at the $A$-site for both Nb- and Ta-based compounds. For $B$-site variation, it shows a blue shift from 420  to 424 cm$^{-1}$ in Co-based systems and from 407  to 414 cm$^{-1}$ in Mn-based systems. Similarly, the M(2) mode in Mn-based systems shifts to higher frequency (440 → 450 cm$^{-1}$) when Nb is replaced by Ta. Three high-frequency modes, A$_{1g}$(3), A$_{1g}$(4), and A$_{1g}$(5), associated with AO$_6$/BO$_6$ octahedral vibrations, generally exhibit a red shift when Co is replaced by Mn at the $A$-site, except for A$_{1g}$(4) in Nb-based compounds, which remains nearly unchanged. These modes display a blue shift upon $B$-site variation from Nb to Ta. Thus, the frequency changes associated with $A$-site variation are consistent with lattice parameter modifications induced by ionic radii, whereas $B$-site variation leads to anomalous shifts in mode positions that cannot be explained solely by mass or ionic radius effects. In addition, the E$_g$(10) mode, observed exclusively in Mn-based systems, shifts to lower frequency when Nb is replaced by Ta.

\subsection{Diffuse reflectance spectroscopy}

For a clear understanding of electronic properties and band gap of MNO/MTO and their Co analogues, we have measured diffuse reflective spectra at ambient conditions. The collected reflectance spectra are converted to their equivalent absorption spectra using Kubelka-Munk function F(R), defined as \cite{Makula}:

\begin{equation}
		F(R) = \frac{K}{S} = \frac{(1-R)^2}{2R},
\end{equation}

where $K$ and $S$ represent absorption and scattering coefficient, respectively; and $R$ is the reflectance of the  sample relative to that of an optically thick standard reference. Finally band gap is estimated using the Tauc method, expressed as \cite{Makula}: 

\begin{equation}
(F(R)\times h\nu)^{\frac{1}{\gamma}} = B_t(h \nu - E_{gap}),
\end{equation}

where $h$ is Planck’s constant, $\nu$ is the photon frequency, $B_t$ is a proportionality constant, and $\gamma$ denotes the nature of the electronic transition ( $\gamma$ = 1/2 for direct and $\gamma$ = 2 for indirect transitions). In this approach, the band gap is obtained from the Tauc plot of (F(R) $\times$ $h\nu$)$^\frac{1}{\gamma}$ vs. $h\nu$, as shown in Fig. 3. The linear portion of the plot, highlighted by the red lines in Fig. 3, is extrapolated to intersect the energy axis, providing the band gap values. 

\begin{figure*}[ht!]
\includegraphics[width=18cm]{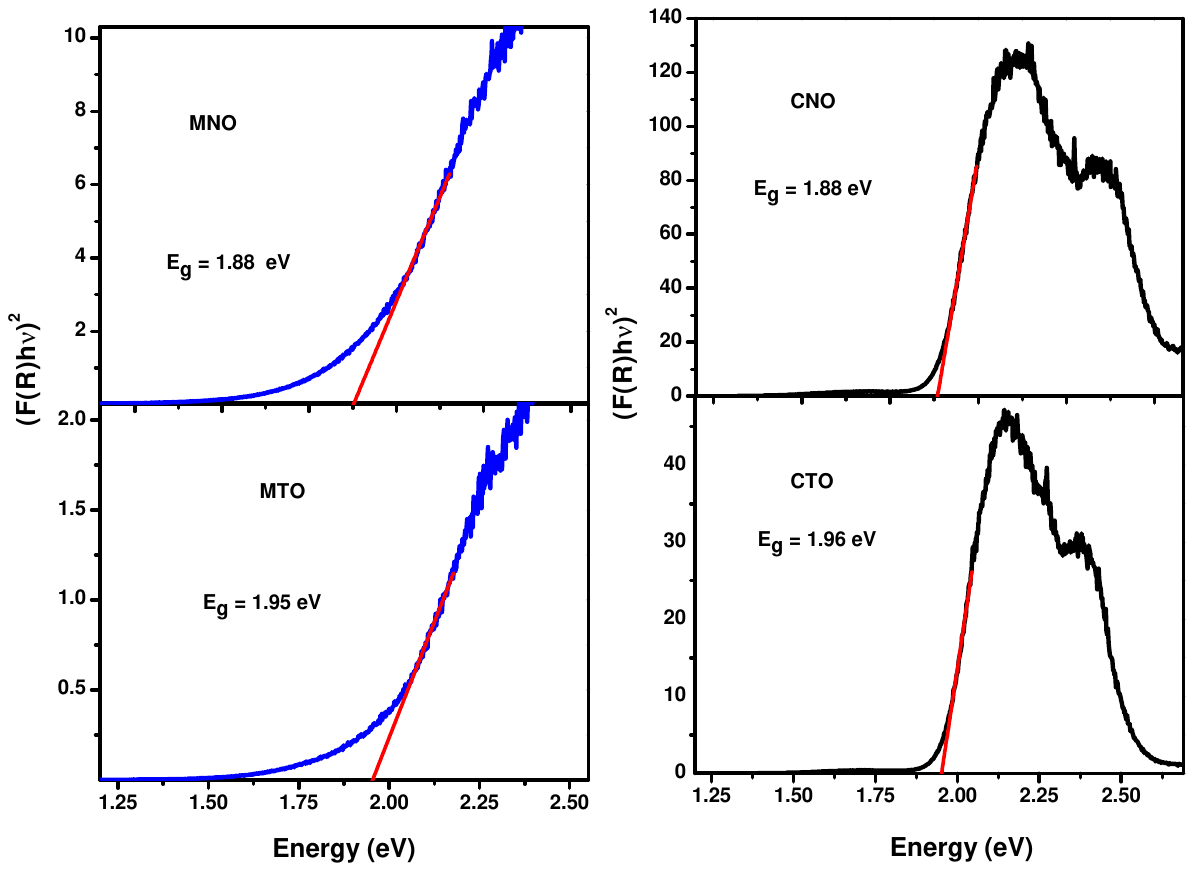}
\caption{\label{fig3} Tauc plots of the four systems derived from diffuse reflectance spectra using Eq.~(1): (a) MNO, (b) MTO, (c) CNO, and (d) CTO. Red solid lines represent the linear fits according to Eq.~(2), and their extrapolation to the energy axis yields the corresponding optical band gaps. } 
\end{figure*}

From this analysis, the estimated band gaps are  $E_{gap}$ = 1.88 and  1.95 eV for MNO and MTO, respectively, indicating a slightly larger gap for the Ta analogue. The corresponding values for the Co-based systems, CNO and CTO, are 1.88 and 1.96 eV, suggesting that all four compounds possess nearly similar band gap energies. However, distinct sub-band-gap absorption features appear in the Co-based systems at approximately 2.1 and 2.34 eV for CNO and at 2.14 and 2.37 eV for CTO, which have been also reported for CNO and attributed to intra- and inter-band electronic transitions in a previous study \cite{Mohanty}. The absence of these features in the Mn-based analogues indicates differences in their electronic structures despite comparable optical gaps, which may also account for the contrasting Raman spectral characteristics observed between the two sub-families.

\subsection{NMR study on MNO and CNO}
We have conducted nuclear magnetic resonance (NMR) measurements on MNO and CNO at room temperature under a magnetic field of 9.3 T to probe the magnetic spin dynamics and local structural environment, which may correlate with the observed Raman mode behavior. Owing to the large nuclear quadrupole moment of $^{183}$Ta and the resulting strong quadrupolar coupling in low-symmetry environments, the corresponding powder spectra are expected to be extremely broadened and were not experimentally resolvable under our measurement conditions \cite{Lapina}. Figure 4 shows the frequency-swept $^{93}$Nb NMR spectra of MNO and CNO. For the spin-$9/2$ nucleus $^{93}$Nb, nine allowed transitions are expected, consisting of a central $-1/2 \leftrightarrow +1/2$ line and four pairs of quadrupolar satellites arising from the finite electric field gradient at the Nb site. In both compounds, all eight satellite transitions are reasonably well resolved together with the central line. The central transition is shifted by approximately 739~kHz in MNO and 846~kHz in CNO relative to the K[NbCl$_6$] reference.

For a half-integer quadrupolar nucleus such as $^{93}$Nb ($I = 9/2$), the central transition (CT) acquires a second-order quadrupolar shift. Using the isotropic powder-average expression for $\eta \approx 0$ \cite{Man},

\begin{equation}
\Delta\nu^{(2)}_{\mathrm{CT}} = -\frac{3\nu_Q^2}{10\nu_L}\left[I(I+1)-\frac{3}{4}\right],
\end{equation}
and $\nu_L \approx 96.9$~MHz at 9.3~T, we obtain $\Delta\nu^{(2)}_{\mathrm{CT}} \approx -56$~kHz for MNO ($\nu_Q \approx 0.87$~MHz) and $\approx -38$~kHz for CNO ($\nu_Q \approx 0.72$~MHz). These second-order shifts are an order of magnitude smaller than the measured CT offsets (739 and 846 kHz), implying that the observed shifts are dominated by paramagnetic hyperfine contributions rather than by quadrupolar second-order effects. The larger paramagnetic shift in CNO suggests a stronger hyperfine interaction ($K$ $\sim$ $A_\mathrm{hf}\chi_\mathrm{loc}$) compared to MNO \cite{Bastow, Baek}, consistent with the higher room-temperature magnetic susceptibility presented in Table III.

		\begin{figure*}[ht!]
\includegraphics[width=16cm]{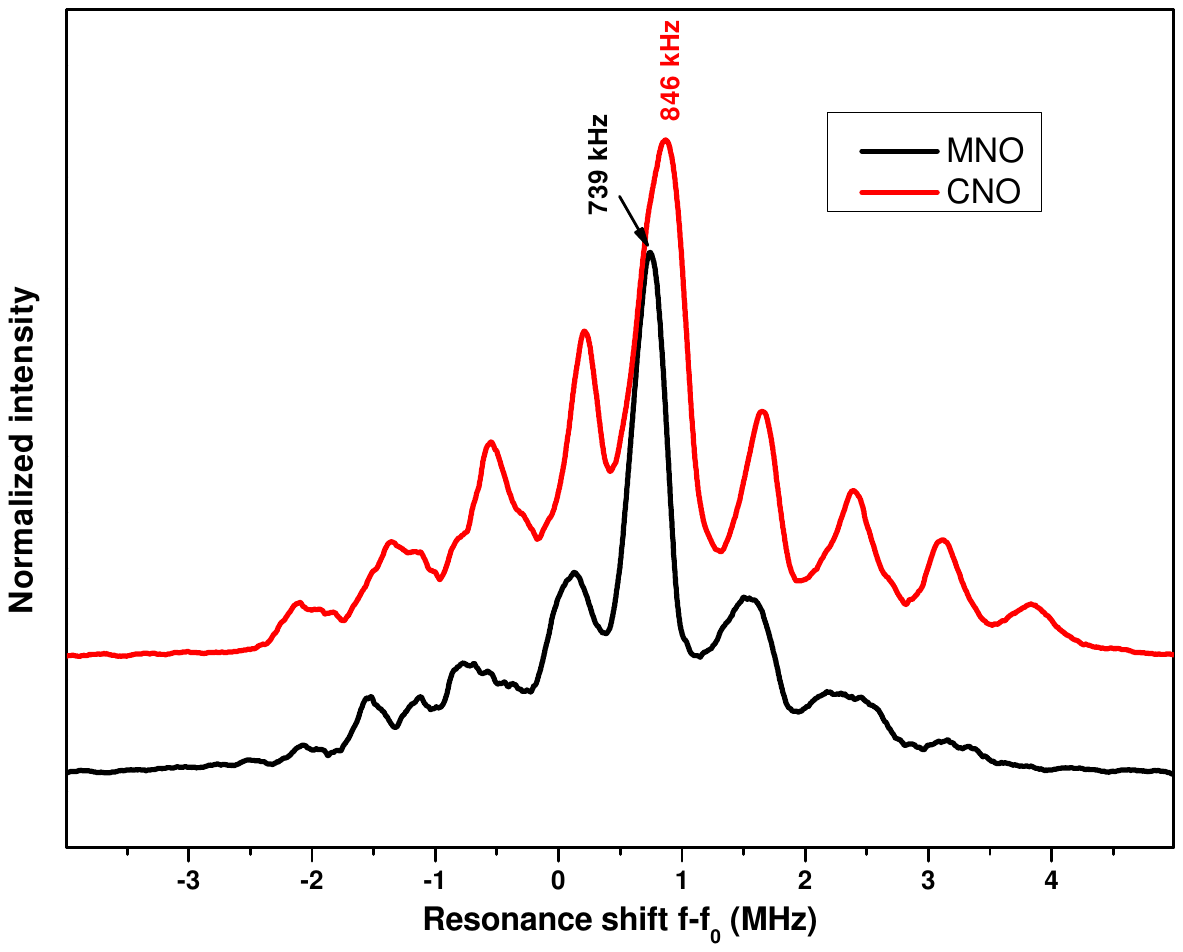}
\caption{\label{fig4} Frequency-sweep NMR spectra of MNO and CNO measured under an applied magnetic field of 9.3 T. The black and red arrows indicate the paramagnetic shifts (f-f$_0$) of MNO and CNO, respectively, determined relative to the nonmagnetic reference K[NbCl$_6$].} 
\end{figure*}

The quadrupole frequencies, extracted directly from the satellite separations, amount to $\nu_Q = 0.87$~MHz for MNO and $\nu_Q = 0.72$~MHz for CNO. For a quadrupolar nucleus, $\nu_Q$ is proportional to the product of the nuclear quadrupole moment $Q$ and the electric field gradient (EFG) at the nuclear site. Since $Q$ is an intrinsic property of $^{93}$Nb and identical in both systems, the larger $\nu_Q$ observed in MNO directly reflects a stronger EFG at the Nb site. This indicates greater deviation from cubic symmetry in MNO than in CNO. The enhanced local distortion inferred from NMR is consistent with the larger number of Raman-active modes observed in the Mn-based compound, supporting a more strongly distorted local structural environment.

In Fig. 5(a)-(b), the transverse magnetization, M$_z$(t) for CNO and MNO, is shown as a function of recovery time (1 $\mu$s–100 ms), recorded using a solid spin-echo excitation pulse sequence. To extract the spin–lattice relaxation time (T1), the magnetization recovery data are fitted using the saturation recovery function, expressed as

\begin{figure*}[ht!]
\includegraphics[width=20cm]{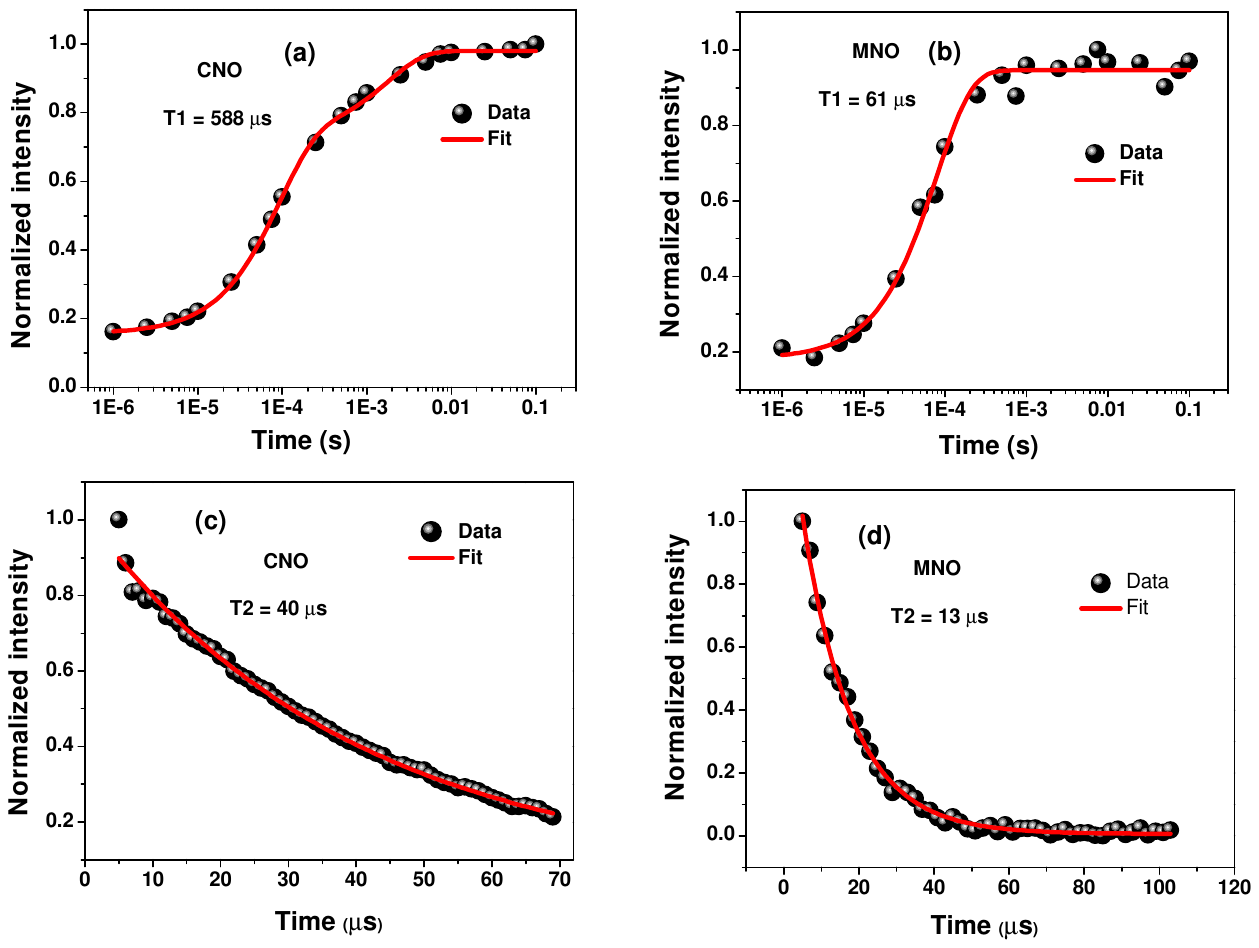}
\caption{\label{fig5} Saturation-recovery measurements for (a) CNO and; (b) MNO. Magnetization-decay measurements for (c) CNO; and (d) MNO. Black circles represent the experimental data, while red lines denote fits using Eqs. (4) and (5), from which the spin–lattice (T1 )  and spin–spin (T2) relaxation constants are determined.} 
\end{figure*}
		
\begin{equation}
	M_z(t) = 1 - B_R\left[C_R \times exp\left({\frac{-t}{t_{11}}}\right) + (1-C_R) \times exp\left(\frac{-t}{t_{12}}\right)\right],
\end{equation}

where $B_R$ and $C_R$ are constants, and t$_{11}$ and t$_{12}$ represent the two characteristic relaxation components contributing to the overall T1 with respective weight factors $C_R$ and (1 – $C_R$). From the best fits, the extracted relaxation times are T1 = 61 $\mu$s for MNO and T1 = 588 $\mu$s for CNO, respectively.

The nearly one order of magnitude shorter T1 in MNO indicates much faster spin–lattice relaxation, suggesting enhanced magnetic fluctuation rates and stronger hyperfine field variations at the Nb site. This behavior reflects a more dynamic local magnetic environment in MNO, consistent with its higher degree of structural distortion inferred from Raman and quadrupolar frequency analyses. The longer T1 in CNO, by contrast, points to relatively weaker spin dynamics and a more uniform local field distribution, in agreement with its comparatively lower structural distortion.

The longitudinal magnetization, $M_{xy}$(t), measured after different delay times, is presented in Fig. 5(c)-(d), showing a gradual decay with time for both systems. The spin–spin relaxation time (T2) is determined by fitting the $M_{xy}$(t)-t data to a single-exponential decay function:

\begin{equation}
M_{xy}(t) = y_0 + exp\left(\frac{-t}{T2}\right), 
\end{equation}

where y$_0$ denotes the offset corresponding to M$_{xy}$ at t = 0, and T2 represents the characteristic time at which the transverse magnetization decays to 37\% of its initial value. The best fits yield T2 = 13 $\mu$s for MNO and T2 = 40 $\mu$s for CNO, respectively.
The much shorter T2 observed for MNO indicates stronger local magnetic field inhomogeneity and enhanced spin–spin interactions, align with greater structural distortion and faster spin dynamics inferred from T1 analysis. In contrast, the longer T2 in CNO reflects a relatively more homogeneous local magnetic environment with slower dephasing of nuclear spins.

\subsection{Low-temperature magnetic investigation}	
			 
Figure 6 presents the temperature-dependent DC magnetic susceptibility $\chi$(T) and inverse of it $\chi^{-1}$(T) for both MNO and MTO. Both materials exhibit an anomalous downturn in their zero-field-cooled (ZFC) and field-cooled (FC) magnetization curves around 120 K and 110 K, respectively, corresponding to their antiferromagnetic (AFM) ordering temperatures (T).   Upon further cooling, $\chi$(T) for MNO shows a pronounced upturn below  $\sim$58 K, with the ZFC curve displaying a distinct kink around 43 K, where a clear divergence from the FC curve appears. A similar kink is observed in MTO around the same temperature, accompanied by a strong ZFC–FC splitting and enhancement of FC magnetization. This low-temperature feature has previously been attributed to the emergence of a weak ferromagnetic (WFM) phase in MTO arising from a canted spin structure \cite{Panja}. Meanwhile, Yu \textit{et al.} \cite{Yu} reported a second anomaly in Co-substituted MNO associated with a spin reorientation transition. Therefore, the secondary anomaly observed in both systems can reasonably be ascribed to either weak ferromagnetic ordering or a spin reorientation transition.

			\begin{figure*}[ht!]
\includegraphics[width=18cm]{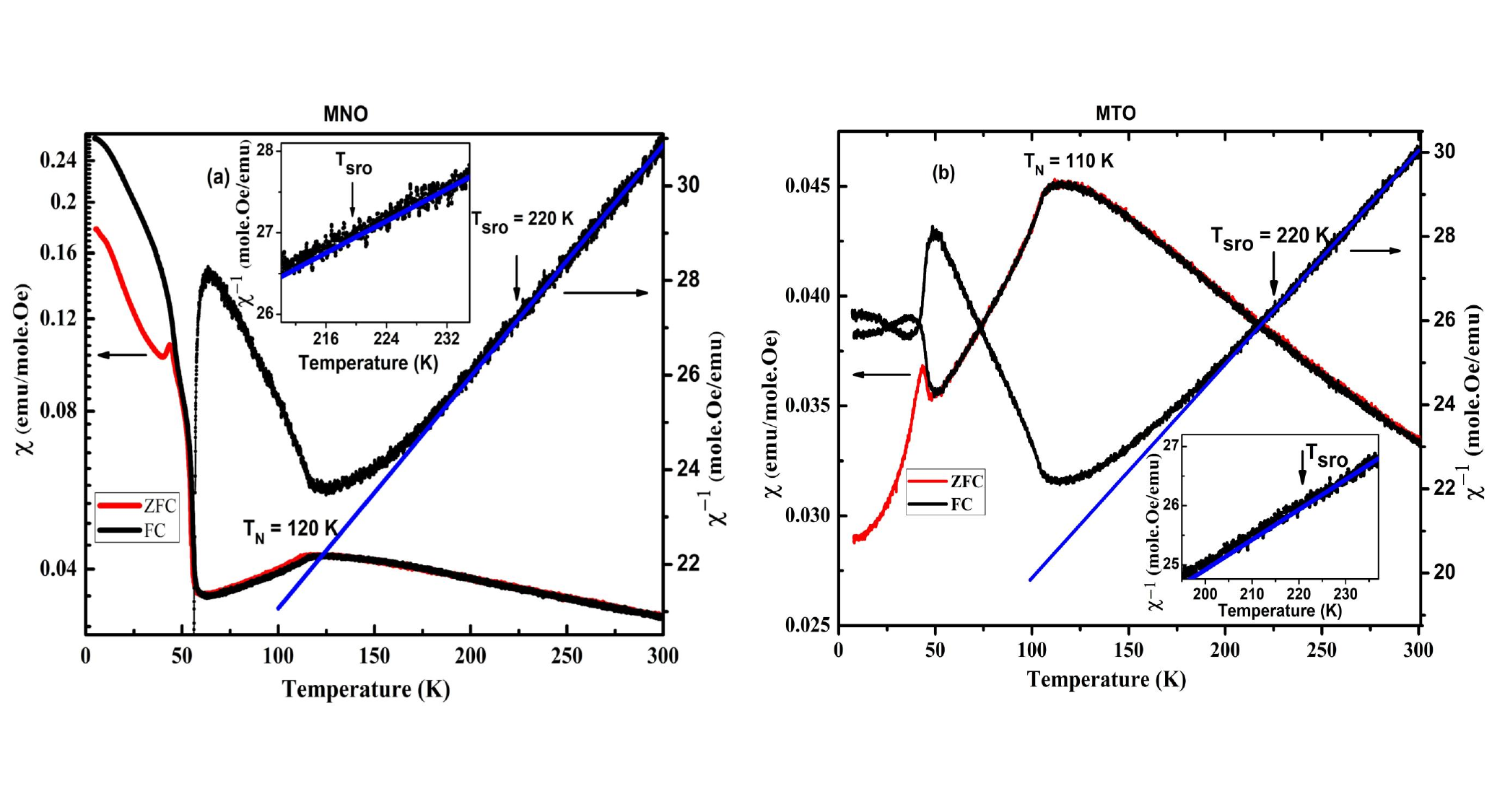}
\caption{\label{fig6} Temperature-dependent magnetic susceptibility $\chi(T)$ and inverse of susceptibility $\chi^{-1}(T)$ for (a) MNO and (b) MTO, measured under zero-field-cooled (ZFC) and field-cooled (FC) protocols. The blue line represents the Curie–Weiss fit to $\chi^{-1}(T)$, which begins to deviate around 220 K, as indicated by the black arrows. Insets in (a) and (b) show enlarged views of $\chi^{-1}(T)$ near the short-range ordering temperature $T_{\mathrm{sro}}$.} 
\end{figure*}
		
			The inverse susceptibility is fitted to the Curie–Weiss (CW) law,
			
	\begin{equation}
		\chi(T) = \frac{C}{(T-\theta_P)},
\end{equation}
		
where  $C$ = $N_A\mu_{eff}^2$/3$k_B$, $N_A$ is Avogadro's number, and  $k_B$ is the Boltzmann constant. The linear CW behavior begins to deviate around 220 K for both systems (inset of Fig. 6(a)-(b)), indicating the onset of short-range magnetic correlations well above $T_N$, similar to the behavior reported for CNO and CTO \cite{Jana2, Jana3}. These results establish that all four compounds evolve from a paramagnetic state to long-range AFM order through an intermediate regime dominated by short-range spin correlations.
							
The extracted parameters, $T_N$, $\mu_{eff}$, $\theta_P$, and $T_{sro}$ for MNO, MTO, along with CNO \cite{Jana2}, and CTO \cite{Jana3} are summarized in Table III. The negative $\theta_P$ values for all systems confirm the dominance of AFM interactions. Moreover, $\theta_P$ increases in the order CTO → CNO → MTO → MNO, correlating with the corresponding increase in $T_{N}$. Notably, the AFM ordering temperatures of MNO and MTO determined in this study are slightly higher than those reported previously, likely due to the high crystalline quality and enhanced homogeneity of the present samples. Interestingly, the frustration index, $f$ = $\left|\frac{\theta_P}{T_N}\right|$, is found to be larger for the Mn-based systems than for their Co analogues, suggesting comparatively stronger competing exchange interactions in the Mn sublattice.

\begin{table*}[htb]
\caption{\label{tab:table2
}
Comparison of  magnetic propertes of MNO and MTO along with Co analogues \cite{Jana2, Jana3}.}
\begin{ruledtabular}
\begin{tabular}{cccccccc}
 Material  & Spin & $ \mu_{eff}$ ($\mu_B$) (predicted)  & T$_N$ (K) & $\mu_{eff}$ ($\mu_B$) & $\chi$ (emu/mole.Oe) & $\theta_P$ (K) & $f$ = $\left|\frac{\theta_P}{T_N}\right|$ \\
\hline
 CNO &  3/2 (Co2+) & 3.87 & 28.5 & 4.67 & 0.0402 & -51.6 & 1.81 \\

 CTO &  3/2 (Co2+) & 3.87 & 21.5 & 4.95 & 0.0439 & -38.5 & 1.79 \\

 MNO &  5/2 (Mn2+) & 5.92 & 120  & 6.41 & 0.0326 & -330 & 2.75 \\

 MTO &  5/2 (Mn2+) & 5.92 & 110  & 6.33 & 0.0336 & -290 & 2.63 \\

\end{tabular}
\end{ruledtabular}
\end{table*}

\subsection{Low temperature Raman study on MNO}

Figure 7 displays the low-temperature Raman spectra of MNO measured over the range of 298 K to 77 K at different temperatures. Upon lowering the temperature, a general blue shift of the Raman modes is observed, which can be attributed to thermal lattice contraction. In addition, the Raman peaks become sharper and more intense at lower temperatures, reflecting reduced phonon–phonon scattering and enhanced vibrational coherence. At 248 K, a clear separation of the E$_g$(8) and E$_g$(10) modes is detected, as indicated by red arrows in Fig. 7(a). These modes, which overlap with the neighboring E$_g$(7) and A$_{1g}$(4) modes at higher temperatures, become distinctly distinguishable upon cooling. Such behavior reflects enhanced lattice distortion, which lifts the mode overlap and allows the intrinsic vibrational features to emerge more clearly. Interestingly, upon further cooling to 148 K, an additional weak octahedral mode emerges around 665 cm$^{-1}$, marked as $\omega$(1) in Fig. 7(a). The intensity of this mode slightly increases with decreasing temperature. Another weak mode, labeled as $\Omega$(1) , appears near 284 cm$^{-1}$ at 77 K. 

\begin{figure*}[ht!]
\includegraphics[width=16cm]{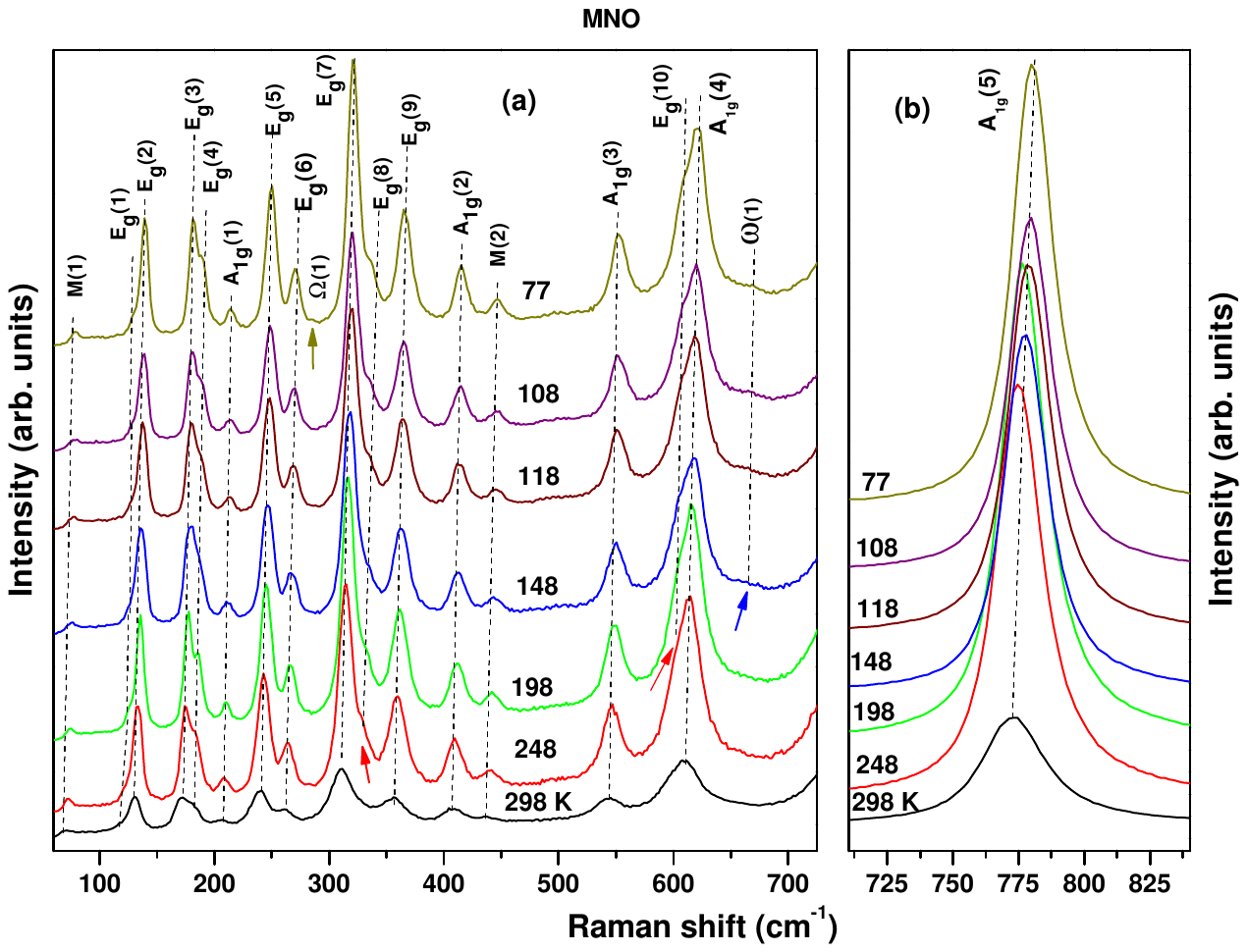}
\caption{\label{fig7} Low-temperature Raman spectra of MNO at selected temperatures between 77 and 300 K over the wavenumber ranges (a) 60–725 cm$^{-1}$ and (b) 710–840 cm$^{-1}$. The appearance or separation of Raman modes is highlighted by arrows, and the black dashed lines trace the temperature evolution of each mode.} 
\end{figure*}

The $\omega$(1) mode has also been observed in our previous studies on CNO and CTO under high pressure, where it was associated with a trigonal-to-monoclinic structural transition.  Therefore, the emergence of the $\omega$(1) mode around 148 K may signify the onset of a monoclinic distortion in MNO. In contrast, the low-frequency $\Omega$(1) mode, which appears within the magnetically ordered state, is likely of magnetic origin. 		

To examine the evolution of Raman modes across $T_{sro}$ and $T_N$ at low temperatures, the spectral profiles are fitted using Lorentzian functions. The resulting mode positions and linewidths (full width at half maximum, FWHM) are shown in Figs. 8 and 9. For a more detailed understanding, the temperature dependence of both parameters is further analyzed using a phonon-phonon anharmonic interaction model \cite{Balkanski}, as described by

		\begin{figure*}[ht!]
\includegraphics[width=18cm]{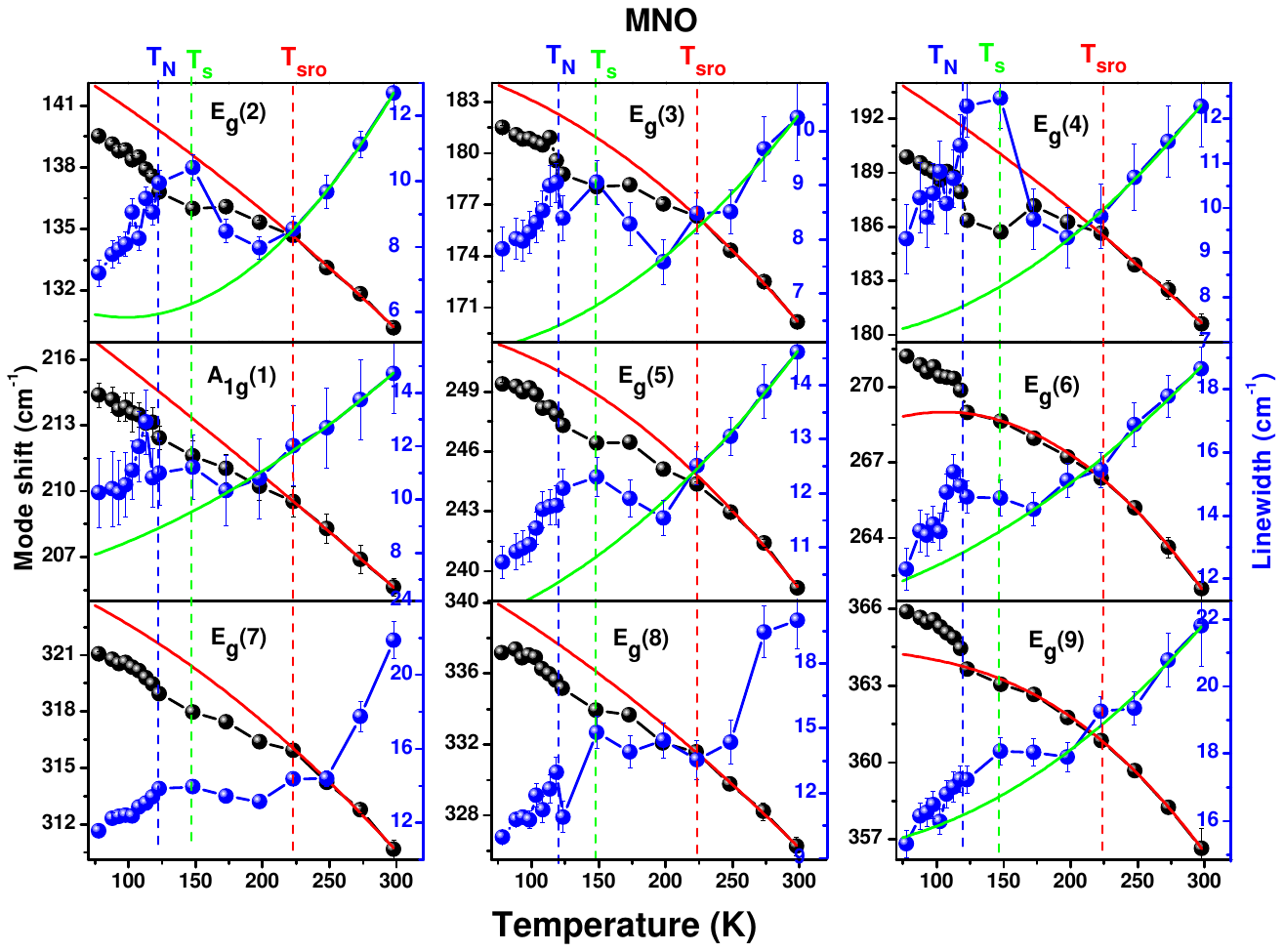}
\caption{\label{fig1} Temperature evolution of the mode frequency and linewidth of the Raman modes of MNO below 400 cm$^{-1}$. The red and green curves represent fits to the experimental mode frequency and linewidth, respectively, as defined in Eqs. (7) and (8). The red and blue dashed lines indicate the short-range ordering ($T_{sro}$) and long-range ordering (T$_N$) temperatures, while the green dashed line marks the possible structural transition temperature.} 
\end{figure*}

\begin{equation}
		\omega(T) = \omega_0+C_1\left[1+\frac{2}{(e^{\hbar\omega_0/2k_BT}-1)}\right]+D_1\left[1+\frac{3}{(e^{\hbar\omega_0/3k_BT}-1)}+\frac{3}{(e^{\hbar\omega_0/3k_BT}-1)^2}\right],
\end{equation}

\begin{equation}
		\Gamma(T) = \Gamma_0+C_2\left[1+\frac{2}{(e^{\hbar\omega_0/2k_BT}-1)}\right]+ D_2\left[1+\frac{3}{(e^{\hbar\omega_0/3k_BT}-1)}+ \frac{3}{(e^{\hbar\omega_0/3k_BT}-1)^2}\right],
\end{equation}

where $\omega_0$ and $\Gamma_0$ represent phonon frequecy and linewidth at zero temperature; $C_1$ ($D_1$), and $C_2$ ($D_2$) are the constants for cubic and quartic anharmonic phonon-phonon scattering process. $\hbar$ is the reduced planck constant and k$_B$ denotes Boltzmann’s constant.

The mode frequencies of several Raman-active phonons, including E$_g$(2), E$_g$(3), E$_g$(4), A$_{1g}$(1), E$_g$(5), E$_g$(7), E$_g$(10), and A$_{1g}$(4), begin to soften significantly  from the expected phonon–phonon anharmonic scattering behavior around 223 K (Figs. 8 and 9), which coincides with the temperature ($T_{sro}$) where magnetic short-range correlations emerge. In addition, the E$_g$(8), and A$_{1g}$(3) modes exhibit modest softening below 223 K relative to extrapolated anharmonic trend. The A$_{1g}$(2) and M(2) modes follow the anharmonic trend down to approximately 200 K, below which A$_{1g}$(2) shows softening, whereas M(2) hardens relative to the anharmonic fit. Upon the onset of long-range antiferromagnetic order below 120 K, all Raman modes, except M(2), display pronounced hardening compared to their frequency evolution in the short-range correlated region (223–123 K). Notably, the E$_g$(6), E$_g$(9), and A$_{1g}$(5) modes remain largely unaffected by short-range magnetic correlations but exhibit strong hardening below $T_N$. In contrast, the M(2) mode shows a distinct softening trend below $T_N$.

\begin{figure*}[ht!]
\includegraphics[width=14cm]{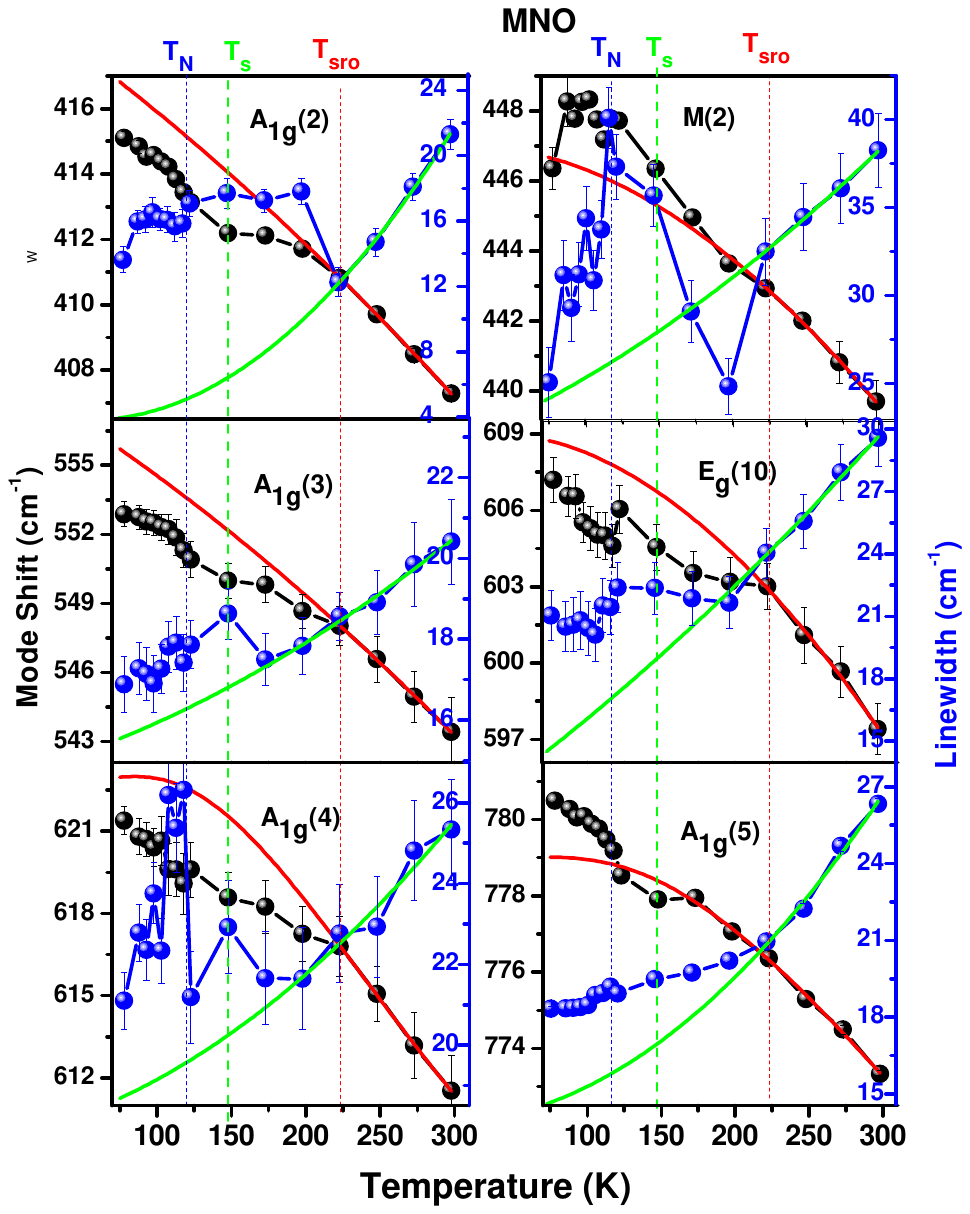}
\caption{\label{fig9} Temperature variation of mode shift and linewidth of higher frequency (400-800 cm$^{-1}$) Raman modes of MNO and their fit red and green curves, respenctively as described in Eqs. (7) and(8), respectively. The red and blue dashed lines indicate the temperatures correspond to short-range ordering ($T_{sro}$) and long-range ordering ($T_N$) temperatures, while the green dashed line marked the possible structural transition temperature.}
\end{figure*}

The largest deviation due to short-range magnetic correlations is observed for the E$_g$(4) mode, which softens by about -5 cm$^{-1}$ around 123 K, just above $T_N$, and by -3.9 cm$^{-1}$ at the lowest measured temperature (77 K) in the long-range magnetic ordered state. Other modes also show appreciable deviations in their frequency evolution at 123 K: E$_g$(2) $\sim$ -2.9, E$_g$(3) $\sim$ -3.4, E$_g$(4) $\sim$ -2.5, E$_g$(5) $\sim$ -2.0, E$_g$(7) $\sim$ -2.7, A$_{1g}$(1) $\sim$ -2.0, A$_{1g}$(3) $\sim$ -2.5, and A$_{1g}$(4) $\sim$ -3.0 cm$^{-1}$. At the lowest temperature, the corresponding deviations are E$_g$(2) $\sim$ -2.3, E$_g$(3) $\sim$ -2.4, A$_{1g}$(1) $\sim$ -2.3, E$_g$(6) $\sim$ +2.3, E$_g$(7) $\sim$ -2.5, E$_g$(9) $\sim$ +1.9, A$_{1g}$(3) $\sim$ -3.0, A$_{1g}$(4) $\sim$ -1.5, and A$_{1g}$(5) $\sim$ +1.5 cm$^{-1}$. Here, negative and positive signs denote phonon softening and hardening, respectively, relative to the anharmonic fit.

In addition to the mode frequency shifts, the Raman mode linewidths exhibit pronounced renormalization in MNO around 200-223 K, reflecting the influence of short-range magnetic interactions. The linewidths of the E$_g$(2), E$_g$(3), E$_g$(4), E$_g$(5), and A$_{1g}$(4) modes follow the Balkanski phonon–phonon scattering model (Eq. 8) down to ~200 K, below which they show an unusual broadening that persists down to ~148 K (Figs. 8 and 9 This anomalous increase signifies enhanced phonon damping due to the onset of spin–phonon coupling associated with short-range magnetic correlations. Upon entering the long-range ordered state below $T_N$, the linewidths of these modes begin to decrease, indicating suppression of dynamic spin fluctuations. 

The A$_{1g}$(1), E$_g$(6), and A$_{1g}$(3) modes also display anomalous linewidth behavior below ~173 K, while the E$_g$(7) and E$_g$(8) modes exhibit distinct and monotonic trends—showing a graduall decrease down to 250 K, a nearly temperature-independent plateau between 250 and 148 K, and a further decrease below $T_N$. The A$_{1g}$(2) and M(2) modes deviate strongly from pure phonon–phonon scattering behavior below ~223 K: the former shows a sudden increase near 200 K followed by a flat region down to 148 K, whereas the latter displays a sharp drop at 200 K and then broadens again down to $T_N$. Below the magnetic ordering temperature, the A$_{1g}$(2) linewidth decreases slightly, while M(2) undergoes a pronounced narrowing. The A$_{1g}$(5) mode also shows substantial deviation from the expected anharmonic trend below 223 K.

Overall, the linewidths of most Raman modes—including E$_g$(2), E$_g$(3), E$_g$(4), E$_g$(5), E$_g$(7), E$_g$(9), A$_{1g}$(3), and A$_{1g}$(5)—are significantly renormalized at $T_N$, followed by a gradual narrowing upon further cooling. In contrast, the A$_{1g}$(4) mode shows a sharp jump in linewidth around $T_N$ before rapidly decreasing. The strong renormalization of both Raman mode frequency and linewidth around $T_{sro}$ and $T_N$ provides clear evidence of strong spin–phonon coupling in MNO, in stark contrast to earlier report \cite{Chen}. Additionally, a secondary anomaly in the frequency and linewidth of several modes appears near 148 K, coinciding with the activation of the octahedral mode $\omega$(2), which likely signifies a magnetostructural transition.

In magnetic materials, several mechanisms contribute to the temperature dependence of phonon mode frequencies, as described \cite{Lockwood, Bhadram, Sohn, Kunwar}: 
		
\begin{equation}
  (\Delta\omega) = (\Delta\omega)_{anh} + (\Delta\omega)_{el-ph} + (\Delta\omega)_{sp-ph},
\end{equation}	

where  $\Delta\omega_{anh}$ represents the phonon–phonon (anharmonic) contribution, which typically yields a smooth, monotonic variation of phonon frequency with temperature. The second term, $\Delta\omega_{el-ph}$, accounts for electron–phonon interactions, which can be neglected for insulating systems such as the present compound due to the absence of free carriers. Consequently, any deviation from the expected anharmonic behavior in phonon frequency can be attributed to the spin–phonon coupling term, $(\Delta\omega)_{sp-ph}$, which arises from modulation of the exchange interaction by lattice vibrations. Thus, the anomalous renormalization of several Raman modes below 223 K is primarily governed by spin–phonon coupling effects \cite{Lockwood, Bhadram, Sohn, Kunwar}

The resultant renormalized phonon frequency in the presence of magnetic interactions can be expressed as \cite{Lockwood, Granado}

\begin{equation}
\omega = \omega_{anh} + \lambda\langle S_i \cdot S_j \rangle,
\end{equation}

where $\omega_{anh}$ represents the phonon frequency in the absence of magnetic influence, $\langle S_i \cdot S_j \rangle$ denotes the spin–spin correlation between neighboring magnetic ions, and $\lambda$ is the spin–phonon coupling constant. The term $\lambda\langle S_i \cdot S_j \rangle$ captures the influence of magnetic exchange on lattice vibrations, and its sign determines whether the phonon mode experiences softening or hardening. A positive $\lambda$ leads to phonon hardening upon the development of antiferromagnetic correlations, whereas a negative $\lambda$ results in phonon softening associated with ferromagnetic interactions. Thus, the magnitude and sign of $\lambda$ are mode-dependent, reflecting the distinct coupling strengths between different phonons and exchange pathways \cite{Lockwood, Lee}. The observation of both hardening and softening behaviors among various Raman modes in MNO thus signifies a non-uniform spin–phonon coupling landscape, arising from competing AFM and FM exchange interactions within the lattice.

To quantify the strength of spin–phonon coupling in MNO, we follow the approach of Lockwood and Cottam \cite{Lockwood, Cottam}, who employed mean-field theory and the two-spin cluster method for antiferromagnetic systems such as MnF$_2$, CoF$_2$, NiF$_2$, and FeF$_2$. In this framework, the spin–spin correlation function is expressed in terms of the short-range order parameter $\Phi(T)$, allowing the phonon frequency shift due to spin–phonon interaction to be written as \cite{Lockwood, Cottam}

\begin{equation}
\Delta \omega = -\lambda S^2 \Phi(T),
\end{equation}

where $\lambda$ is the spin–phonon coupling constant as defined in Eq. 10, and $\Phi(T)$ captures the degree of short-range magnetic order. Below $T_N$, $\Phi(T)$ can be estimated using mean-field theory, while above $T_N$, the two-spin cluster approximation provides a reliable description \cite{Lockwood, Cottam}.

This approach has been extensively used to determine $\lambda$ for numerous magnetic insulator using the formula \cite{Pal, Aytan, Poojitha}:

\begin{equation}
\lambda = -\frac{\omega(T_\mathrm{low}) - \omega_\mathrm{anh}(T_\mathrm{low})}{[ \Phi(T_\mathrm{low}) - \Phi(2T_N) ] S^2},
\end{equation}

where $\omega(T_\mathrm{low})$ is the experimentally observed Raman mode frequency at the lowest measured temperature, and $\omega_\mathrm{anh}(T_\mathrm{low})$ is the corresponding frequency predicted from the anharmonic phonon model (Eq. 7). Using the values of $\Phi(T)$ from Ref. \cite{Cottam}, the spin–phonon coupling constant for MNO at 77 K is found to lie in the range $|\lambda| \sim 0.5$–$1.25$ cm$^{-1}$, comparable to those reported for MnF$_2$ and FeF$_2$ using the same methodology \cite{Lockwood}.

Interestingly, estimating $\lambda$ at 123 K, just above the long-range ordering onset, yields significantly higher values ($|\lambda| \sim 4$–$8$ cm$^{-1}$), reflecting the enhanced influence of spin correlations in the short-range ordered regime.

\subsection{Low temperature Raman study on MTO}

Figure 10 presents the temperature-dependent Raman spectra of MTO measured down to 77 K at selected temperatures. Overall, the Raman modes initially exhibit a blue shift upon cooling, consistent with the expected thermal lattice contraction. In terms of intensity, the Raman spectra exhibit a general enhancement down to 198 K, followed by a gradual decrease at lower temperatures. At 248 K, a weak Raman mode, labeled $\nu$(1), emerges on the right-hand tail of the E$_g$(4) mode near 200 cm$^{-1}$, as highlighted by the navy arrow in Fig. 10(a), suggesting that its activation may be related to local symmetry breaking induced by short-range spin correlations. At the same temperature, a clear separation of the E$_g$(6) mode from E$_g$(5) is detected, indicating enhanced structural distortion. Upon further cooling to 173 K, the previously overlapping E$_g$(9) mode starts to distinctly visible. Interestingly, at 123 K, a weak hump-like feature appears in the 65–90 cm$^{-1}$ region, which can be traced down to 93 K and is labeled as excitation E (purple arrow in Fig. 10(a)). Its appearance near the long-range magnetic ordering temperature (T$_N$) suggests a magnetic origin. Due to the gradual reduction in overall Raman intensity at lower temperatures, this excitation becomes difficult to recognize at 77 K.

\begin{figure*}[ht!]
\includegraphics[width=16cm]{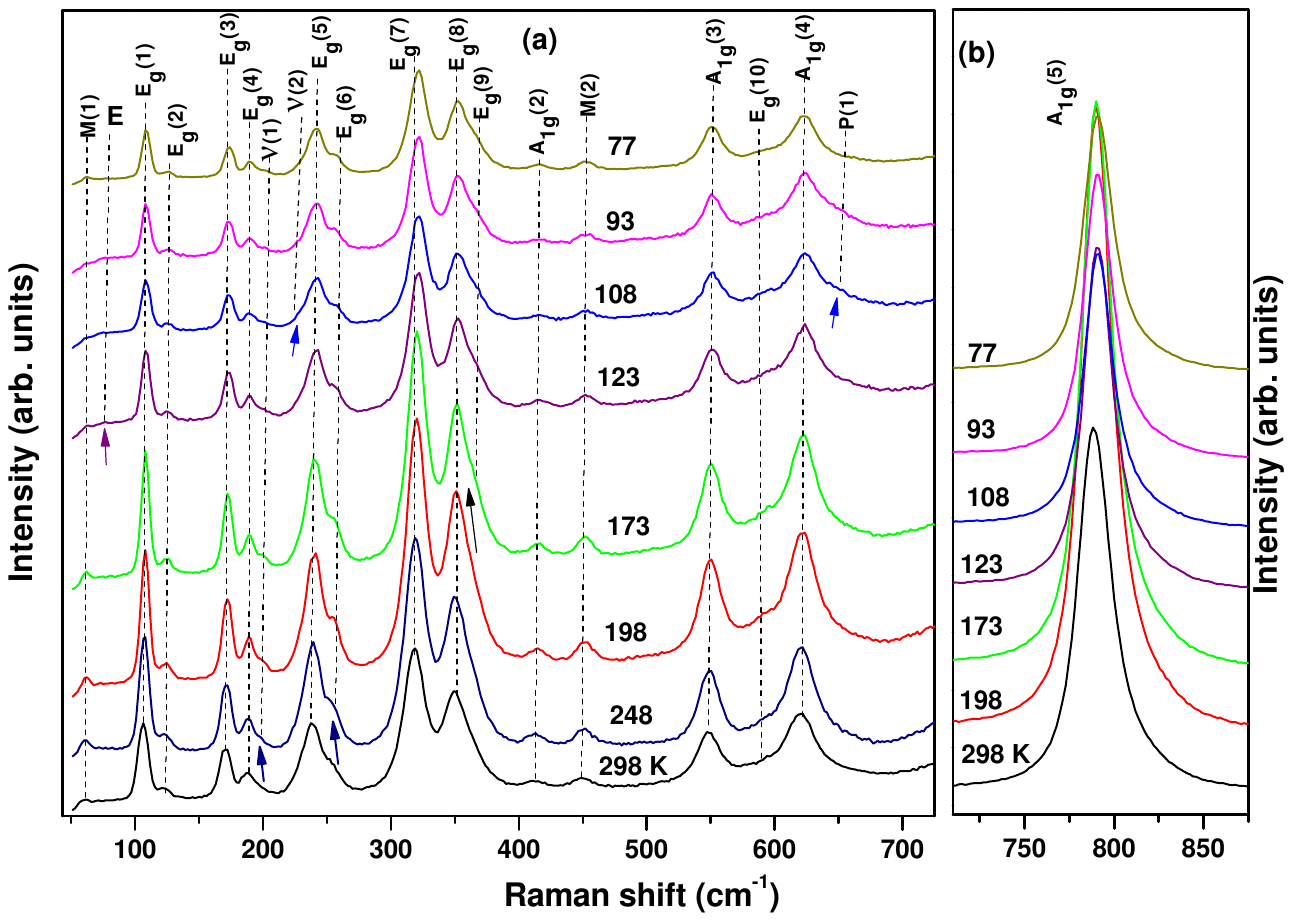}
\caption{\label{fig10} Raman spectra of MTO measured at various temperatures down to 77 K in the frequency ranges (a) 50–725 cm$^{-1}$ and (b) 710–875 cm$^{-1}$. The temperature evolution of each mode is indicated by black dashed lines, and the appearance/seperation of new modes is highlighted by arrows.} 
\end{figure*}

Most notably, two new Raman modes, $\nu$(2) at 225 cm$^{-1}$ and  P(1) at 649 cm$^{-1}$, appear at 108 K, as indicated by the blue arrows in Fig. 10(a). A magnified view of $\nu$(2) at 93 K is shown in Fig. 11(a), while a Lorentzian fit of P(1) alongside neighboring modes is presented in Fig. 11(b).  The appearance of the P(1) modes closely mirrors their activation during the trigonal-to-monoclinic structural transition in high-pressure studies on CNO and CTO \cite{Jana2, Jana3}, suggesting that a similar structural distortion occurs at 108 K, coinciding with $T_N$. On the other hand, the emergence of the weak mode $\nu$(2) in the low-frequency regime could be attributed to magnetic interactions. The P(1) mode persists down to 93 K but becomes challenging to detect at 77 K due to the overall reduction in Raman intensity.

\begin{figure*}[ht!]
\includegraphics[width=12cm]{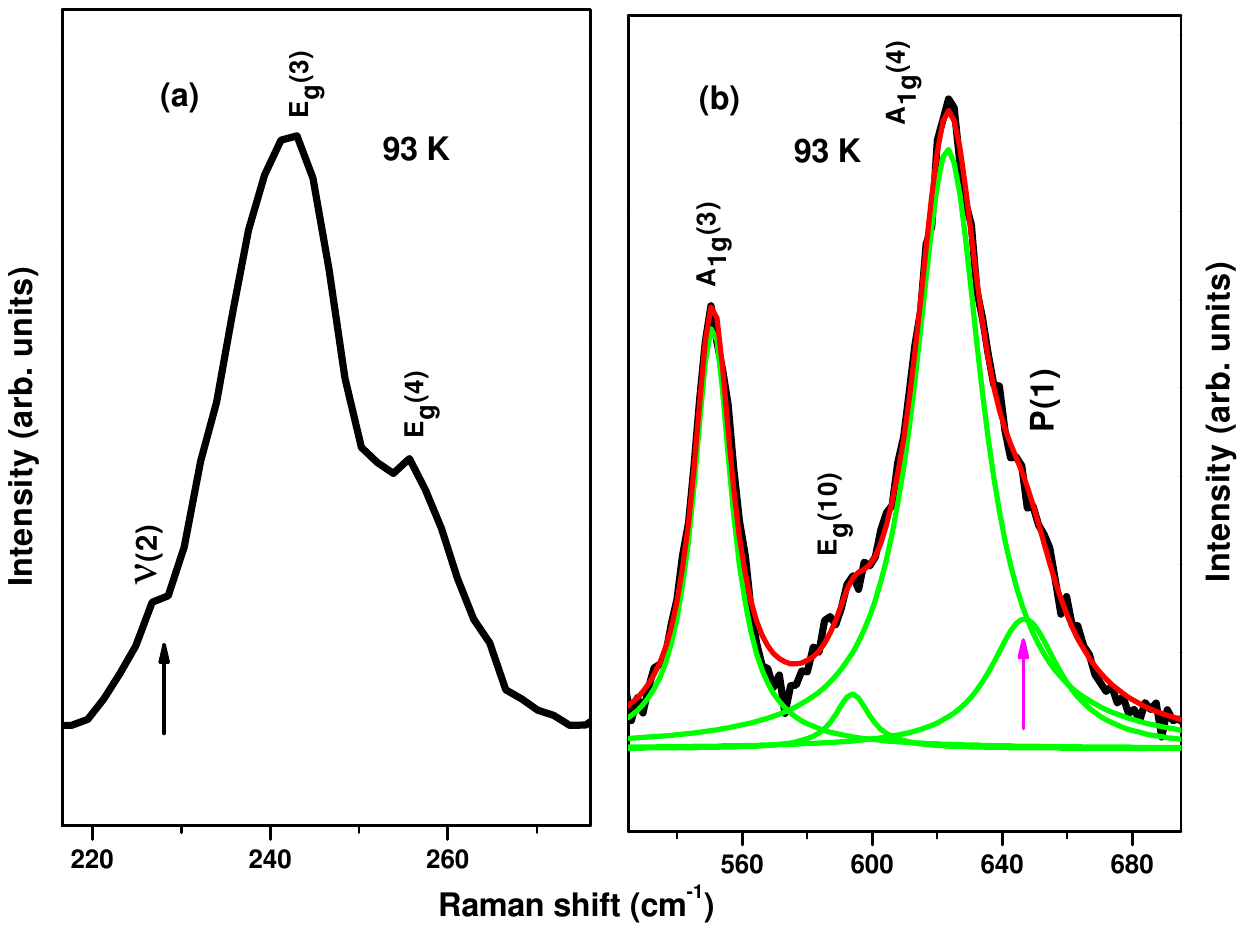}
\caption{\label{fig11} Magnified view of the Raman spectrum at 93 K over the frequency ranges (a) 213–276 cm$^{-1}$ and (b) 525–695 cm$^{-1}$. The emergence of two new modes, $\nu$(2) and P(1), is indicated by black and magenta arrows, respectively.} 
\end{figure*}

The low-temperature evolution of Raman mode frequencies and linewidths of MTO, compared to their anharmonic fits (Eqs. 7 and 8), is shown in Figs. 12 and 13. Upon cooling down to 248–223 K, most Raman modes follow the expected phonon–phonon scattering behavior, with the exception of the linewidths of E$_g$(8) and E$_g$(9) modes. Below 248 K, the frequencies of E$_g$(4) and E$_g$(10) show noticeable softening relative to the anharmonic fit, whereas E$_g$(1) and E$_g$(6) begin to deviate around 223 K. In contrast, E$_g$(3), E$_g$(5), E$_g$(8), A$_{1g}$(2), A$_{1g}$(3), and A$_{1g}$(4) exhibit anomalous hardening below 200 K. The mode frequency of the E$_g$(9) mode starts to hardens at 248 K and it shows highest deviation among all the modes in the short-range spin correlations region. E$_g$(2) shows a clear slope change around 200 K, and A$_{1g}$(5) displays subtle deviations from the anharmonic trend.

\begin{figure*}[ht!]
\includegraphics[width=18cm]{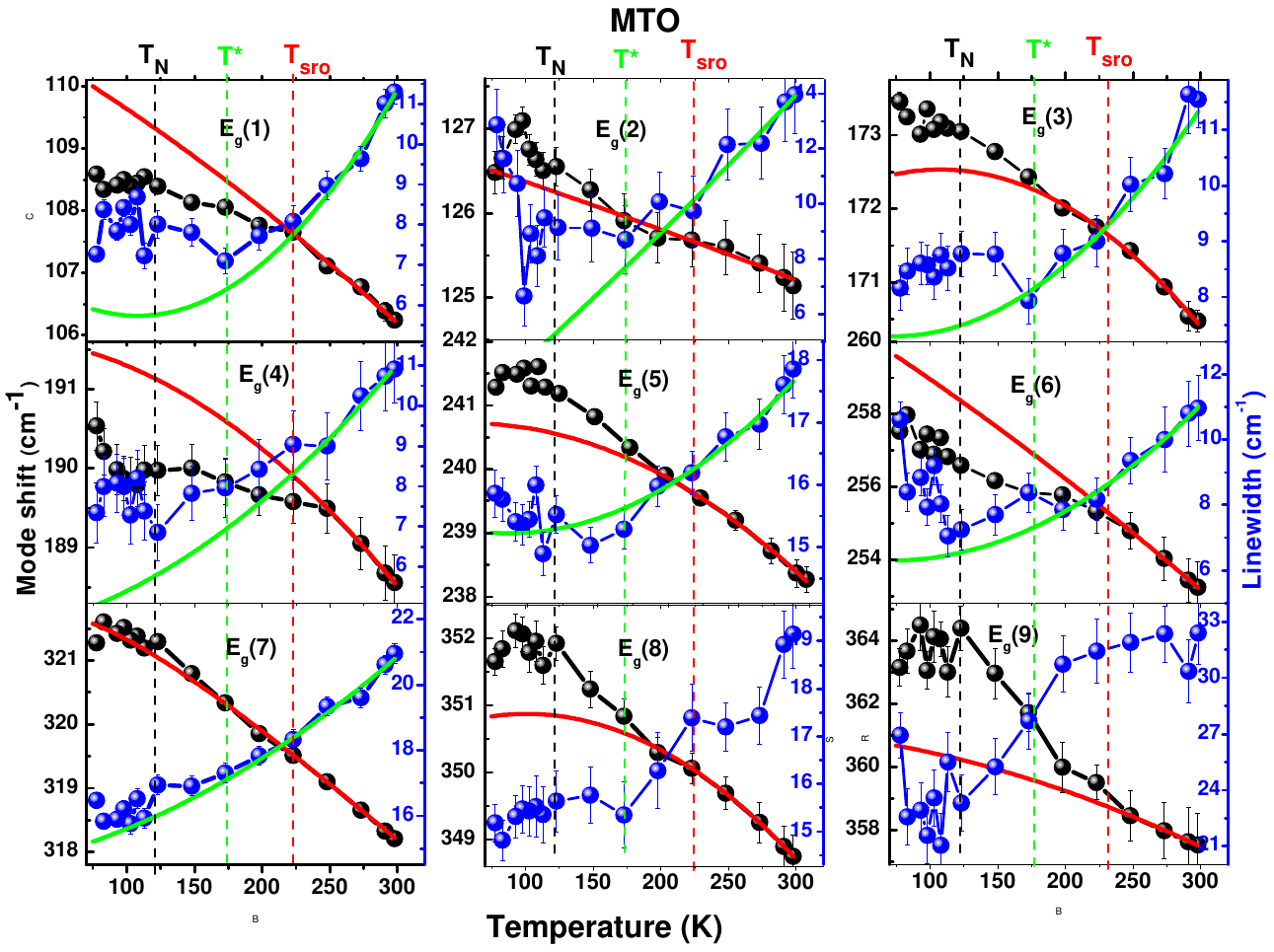}
\caption{\label{fig12} Temperature evolution of Raman mode frequency and linewidth of MTO down to 77 K. The red and green curves represent anharmonic phonon–phonon decay fits, as described in Eqs. (7) and (8), respectively. The red dashed line indicates the magnetic short-range ordering temperature ($T_{sro}$), the black dashed line marks the long-range ordering temperature (T$_N$), and the green dashed line highlights an additional anomaly observed in specific Raman modes around 173 K.} 
\end{figure*}

At the onset of long-range magnetic ordering ($T_N$), pronounced renormalization of nearly all Raman mode frequencies is observed. While E$_g$(1) and A$_{1g}$(3) remain nearly temperature independent in the magnetically ordered state, E$_g$(4) and E$_g$(6) harden significantly compared to their behavior in the short-range correlation regime. In contrast, the $E_g(5)$, $E_g(9)$, $M(2)$, $A_{1g}(2)$, and $A_{1g}(4)$ modes display pronounced softening,, and most other modes show clear deviations from their anharmonic extrapolations below $T_N$. The frequency shifts associated with short-range correlations just above T$_N$ (113 K) are estimated as E$_g$(1) $\sim$ -1, E$_g$(4) $\sim$ -1.2, E$_g$(5) $\sim$ +0.7, E$_g$(6) $\sim$ -1.7, E$_g$(9) $\sim$ +4.3, and A$_{1g}$(2) $\sim$ +1.2 cm$^{-1}$. At the lowet temperature 77 K, well within the long-range ordered phase, the corresponding shifts are E$_g$(1)) $\sim$ -1.2, E$_g$(3) $\sim$ +1, E$_g$(4) $\sim$ -1, E$_g$(6) $\sim$ -2, E$_g$(8) $\sim$ +0.8, E$_g$(9) $\sim$ +2.6, and A$_{1g}$(3) $\sim$ +1.2 cm$^{-1}$. These substantial and mode-dependent frequency renormalizations provide clear evidence of strong spin-phonon coupling and magnetoelastic interactions in MTO. Using Eq.~(12), the spin--phonon coupling constants associated with short-range and long-range magnetic ordering are estimated to be $|\lambda| \sim 1.12$--$6.88$~cm$^{-1}$ at 113~K and $|\lambda| \sim 0.3$--$1$~cm$^{-1}$ at 77~K, respectively. These values are slightly smaller than those obtained for the MNO analogue. 
  
\begin{figure*}[ht!]
\includegraphics[width=14cm]{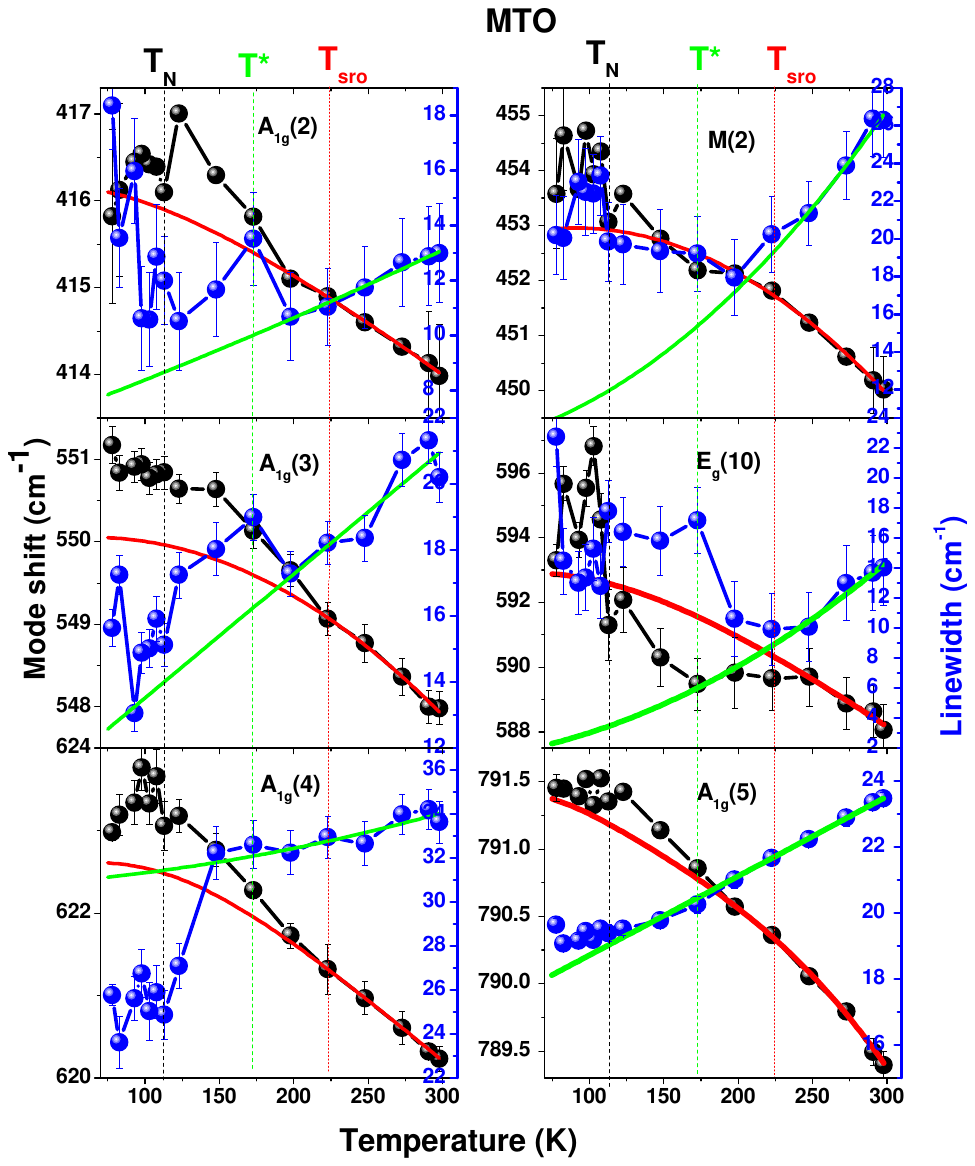}
\caption{\label{fig13} Evolution of high frquency Raman mode shift and linewidth of MTO with temperature down to 77 K. The red and black dashed lines indicate the temperatures correspond to short-range ordering ($T_{sro}$) and long-range ordering ($T_N$) temperatures, while the green dashed line marked an additional anomaly around 175 K.} 
\end{figure*} 

The linewidths of several Raman modes in MTO also exhibit pronounced deviations from standard phonon–phonon anharmonic behavior upon cooling. In particular, E$_g$(1), E$_g$(2), E$_g$(4), and M(2) narrow following Eq. 8 down to  223 K then deviate from the fit and marginally change at lower temperatures down to $T_N$. While, the linewidth of E$_g$(10) mode initially decreases as expected from the anharmonic model, however below 223 K it remarkably broadens down to the long-range magnetic transition temperature accompanied by  a irregularity around 173 K. Additional E$_g$(5), E$_g$(6), A$_{1g}$(2), and A$_{1g}$(3) display  anomalies below 200 K, with a notable peculiarity near 173 K. The linewidth of E$_g$(3) follows the anharmonic trend down to 173 K, below which it increases markedly. E$_g$(8) exhibits strong anomalies near both 223 K and 173 K, while E$_g$(9) shows a gradual decrease down to 200 K followed by a rapid reduction in the 200–123 K range. These observations suggest that the anomalies around 223 K are linked to the emergence of short-range magnetic order, whereas those around 173 K may reflect further modifications in short-range spin correlations or subtle structural distortion.

Upon entering the long-range antiferromagnetic state below $T_N$, drastic changes in linewidth are observed: E$_g$(2), E$_g$(6), and A$_{1g}$(2) exhibit substantial broadening, while E$_g$(4) and E$_g$(5) show a significant upturn. In contrast, E$_g$(3) and E$_g$(8) gradually narrow in the 108–77 K range (Fig. 12 ). The A$_{1g}$(4) mode undergoes a pronounced drop in linewidth around $T_N$, whereas E$_g$(10) initially decreases at $T_N$ but then sharply broadens below 95 K (Fig. 13). These linewidth anomalies reflect strong spin–phonon coupling and magnetoelastic effects, indicating that the lattice dynamics of MTO are intimately linked to both short-range and long-range magnetic ordering.

\subsection{Discussion}

Analysis of the ambient Raman spectra of the four $A$$_4$$B$$_2$O$_9$ compounds demonstrates that the Raman mode positions, intensities, and the number of active modes cannot be explained solely by ionic-radius or atomic-mass considerations. We detect 16–17 Raman-active modes in MTO and MNO, but only 11 and 15 modes in CNO and CTO, respectively. Despite the larger ionic radius of Mn$^{2+}$ and the larger unit cell volume of Mn-based compounds, the Mn systems show more Raman-active modes than the Co analogues, indicating enhanced local distortion in Mn compounds. This observation is further strengthen by the NMR findings with higher quadrupolar frequency and faster nuclear relaxation constants (T1 and T2) in MNO than CNO.  Thus the difference in local strcuture of Mn- and Co-based compounds play domination role in their contrasting Raman spectra. However, Several other notable exceptions particularly in $B$-site variants further point to factors beyond simple mass or size effects that strongly influence the lattice dynamics of these systems.  For instance, additional Raman modes and a general shift of several modes toward higher frequencies are observed in CTO compared to CNO, which has been attributed to spin–orbit coupling (SOC) effects, since Nb$^{5+}$ and Ta$^{5+}$ have nearly identical ionic radii \cite{Jana3}. Interestingly, similar to CTO, the E$_g$(1) mode appears at 106 cm$^{-1}$ in MTO but is extremely weak/absent in the Nb-based counterparts, MNO and CNO. Moreover, the A$_{1g}$(1) mode is detected only in CTO and MNO—i.e., in the Ta compound for Co systems and in the Nb compound for Mn systems—around 205 cm$^{-1}$ with comparable intensity, reflecting the complex interplay between local structural environment and lattice vibrations. The E$_g$(3) and E$_g$(4) modes exhibit sharper resolution in the Nb-based Co system and the Ta-based Mn system, whereas the E$_g$(5) and E$_g$(6) modes are fully resolved only in the Ta–Co and Nb–Mn compounds, respectively. Furthermore, an intensity transfer between the closely spaced E$_g$(8) and E$_g$(9) modes is observed in the Mn systems. Much stronger spectral weight transfer is also evidenced between neighbouring E$_g$(1) and E$_g$(2) modes in both Mn and Co-based analogues when Nb is replaced by Ta.  Generally, isostructural compounds with different ionic radii produces different lattice distortion whcih dictate phonon shift and scattering intensity as in the case RMnO$_3$ (R = rare erath ion) \cite{Laverdière}. Similarly the substitution of the Sm  with larger ionic radius in place of the Gd site of SmCrO$_3$, orthorhombic distortion decreases, which result in the decrease in  the mode intensity \cite{Das}. 
	
Thus contrasting Raman scattering behaviors in the present LME materials suggest that the hybridization among Co/Mn, Nb/Ta, and O orbitals plays crucial role in modulating the Raman-active phonons. Such hybridization likely affects specific phonon modes through variations in local bond lengths and covalency, analogous to the behavior reported in pyrochlore iridates (Sm$_{1-x}$Bi$_x$)$_2$Ir$_2$O$_7$ \cite {Rosalin} and manganates M$_2$Mn$_2$O$_7$ (M = Tl, In) \cite{Brown}. Importantly, this orbital hybridization may also alter the superexchange pathways and anisotropic exchange interactions, thereby influencing the strength of spin–phonon coupling and magnetoelectric correlations at low temperatures in these $A_4B_2$O$_9$ systems. The difference in the orbital hybridization could also be responsible for the slightly higher band gap observed in Ta variants.

Orbital hybridization plays a key role in understanding why replacing Nb (4$d$) with Ta (5$d$) modifies the magnetic ordering temperature $T_N$ and lattice parameters. In triangular-lattice antiferromagnets Sr$_3$Co(Nb,Ta)$_2$O$_9$, the Nb-based compound exhibits higher-temperature antiferromagnetic transitions than its Ta analogue, attributed to differences in the Co–O–Nb–O–Co and Co–O–Ta–O–Co hybridization pathways \cite{Lal}. Similar hybridization-driven magnetic contrasts have been reported in SrLaCu(Sb,Nb)O$_6$ \cite{Watanabe}, CNO and CTO magnetothermal transport \cite{Ueno}, and double perovskites Cs$_2$(Ag,Na)FeCl$_6$, where orbital hybridization strongly influences exchange interactions, $T_N$, and magnetostructural distortions despite identical crystal symmetry \cite{Mopoung, Mopoung2}. Accordingly, the lower $T_N$ in Ta-based compounds compared to their Nb counterparts can be linked to the more spatially extended Ta 5$d$ orbitals, which also produce subtle changes in lattice parameters and the \textit{c/a} ratio despite similar ionic radii of Nb$^{5+}$ and Ta$^{5+}$. In contrast, $A$-site substitution has a stronger structural impact: replacing Co with Fe and Mn systematically increases the unit-cell volume and $T_N$, while reducing the \textit{c/a} ratio, consistent with their increasing ionic radii. A clear correlation emerges across both $A$- and $B$-site substitutions, where decreasing \textit{c/a} enhances $T_N$, highlighting strong spin–lattice–orbital entanglement. Structurally, alternating edge-sharing connectivity between magnetic ($A$) and nonmagnetic ($B$) octahedra in the buckled layers of $A_4B_2$O$_9$ promotes strong $A$–O–$B$–O–$A$ hybridization, enabling efficient tuning of exchange interactions and electronic structure. 

The pursuit of highly efficient multifunctional devices strongly relies on a deep understanding of the coupling among magnetic, vibrational, electric polarization, and structural degrees of freedom. In this context, antiferromagnetic spin-driven magnetoelectric materials stand out, as they are free from stray magnetic fields and enable efficient control of electric (magnetic) degrees of freedom through manipulation of magnetic (electric) properties via lattice interactions. Therefore, the observation and detailed understanding of spin–phonon coupling in such systems not only provide crucial insights into the microscopic origin of magnetoelectric coupling but also hold strong potential for practical applications under epitaxial strain, applied pressure, low temperature, or magnetic field conditions \cite{Sun}. Our present study eshtablish pronounced spin-phonon coupling below 223 K due to the developement of short-range spin correlations and further renormalization of phonon self energy parameters around $T_N$ in both MNO and MTO. The microscopic origin of magnetoelectric coupling has been attributed to the magnetic exchange strcition process for the collinear spin-alligned MNO/MTO systems. In this mechanism, the magnetic exchange interactions driven lattice modulation would modify the phonon frequency and thus the anamolus softening or hardening relative to their expected anharmonic behaviour of mode frequency can be observed. This can well explain the renormalization of stong mode frequency around $T_{sro}$ and $T_N$. However, the Raman mode linewidth, inverse of the phonon life time is not affected by the modification in lattice parameters, and is modulated if additional scattering channels are set in due to other process such as electron-phonon or spin-phonon interactions. In this sense the observed substantial anamolus behaviour in several modes of MNO and MTO indicate the complex role of other factor and involvement of the orbital degrees freedom in inducing spin-phonon and magnetoelectric coupling. For a canted spic-structure system like CNO/CTO, a spin-current model (inverse DM interactions) is most probable to contribute to the spin-phonon coupling. In the case of present systems MNO/MTO, low temperature neutron diffraction studies eshtablished the collinear arragement of magnetic spins without any canting. Howevr, a small canting of spins can not be completely ruled out, considering that a ferromagnetic like magnetization enhancement due to antiferromagnetic spin canting is obsrved well below $T_N$ in both systems. Otherwise, the sinlge-ion anisotropy could play significant role in the renormalization of mode frequency and linewidth. For Mn-based systems with S = 5/2, the orbital angular momentum is expected to be quenched (L=0), and hence no significant orbital contribution to the magnetic moment is anticipated. However, magnetic susceptibility measurements reveal a moderate orbital contribution to the total magnetic moment in both MNO and MTO. This behavior likely originates from higher-order (spin-orbit coulping) orbital mixing induced by local structural distortions and such orbital modifications can give rise to single-ion anisotropy, thereby enhancing spin–phonon coupling and contributing to the  magnetoelectric coupling \cite{Sohn, Toft, Lado}.  
We note that the spin–phonon coupling (SPC) below $T_N$ is reported to be comparatively stronger in Co-based systems than in their Mn counterparts \cite{Park}. Considering spin–orbit coupling (SOC), the magnetic interactions in the $A_4B_2$O$_9$ family can be described by the effective spin Hamiltonian \cite{Xiang, Sannigrahi} 

\begin{equation}
		H_{spin} = \sum_{i<j}{J_{ij}\bold{S}_i.\bold{S}_j} + \sum_{i<j}{\bold{D}_{ij}\bold{S}_i \times \bold{S}_j} + \sum_{i}{A_{i}S^2_iz},
\end{equation}, 

where the first term represents the Heisenberg exchange interaction, the second term accounts for the Dzyaloshinskii–Moriya (DM) interaction associated with spin canting, and the third term describes single-ion anisotropy. The exchange parameters $J_{ij}$, $\mathbf{D}_{ij}$, and $A_i$ depend sensitively on the positions of both magnetic and nonmagnetic ions. Consequently, lattice vibrations can modulate all these coefficients, giving rise to spin–phonon coupling and phonon renormalization in both short- and long-range magnetically ordered states \cite{Sannigrahi}. In Co-based systems, the stronger SOC enhances spin canting and single-ion anisotropy, leading to a more pronounced SPC compared to Mn analogues \cite{Park}. Furthermore, among the Co-based compounds, the Ta variant exhibits stronger SPC than its Nb counterpart, consistent with enhanced SOC arising from the heavier Ta$^{5+}$ ion at the $B$ site \cite{Park, Jana2, Jana3}. In the present Mn-based systems, most Raman modes of MNO soften relative to anharmonic behavior below $T_{sro}$, undergo further renormalization near $T_N$, and subsequently harden upon further cooling, with the exception of the M(2) mode. In contrast, MTO exhibits a more complex response: several modes soften below $T_{sro}$, while others harden relative to the anharmonic trend, followed by further renormalization at $T_N$. This mode-dependent behavior makes it difficult to unambiguously identify which Mn analogue exhibits stronger SPC. The presence of a possible structural transition between $T_{sro}$ and $T_N$ in MNO, and near $T_N$ in MTO, further complicates the interpretation. Consistently, the Raman linewidths also display markedly different anomalous behaviors in the two systems. Nevertheless, our results clearly establish that both MNO and MTO are strongly influenced by the emergence of short-range and long-range magnetic correlations, although in distinct ways, underscoring the crucial role of the nonmagnetic $B$-site cation in tuning spin–lattice interactions. To further probe these effects, we compare the temperature evolution of the integrated intensities of four representative Raman modes in Fig.~14. For both compounds, the intensities initially increase upon cooling down to $T_{sro}$ and then decrease below $\sim$200 K. Strikingly, below $T_N$, MNO shows an overall enhancement of spectral weight, whereas MTO exhibits a reduction, indicating that long-range magnetic ordering redistributes the Raman spectral weight in fundamentally different ways in these two systems.  

\begin{figure*}[ht!]
\includegraphics[width=16cm]{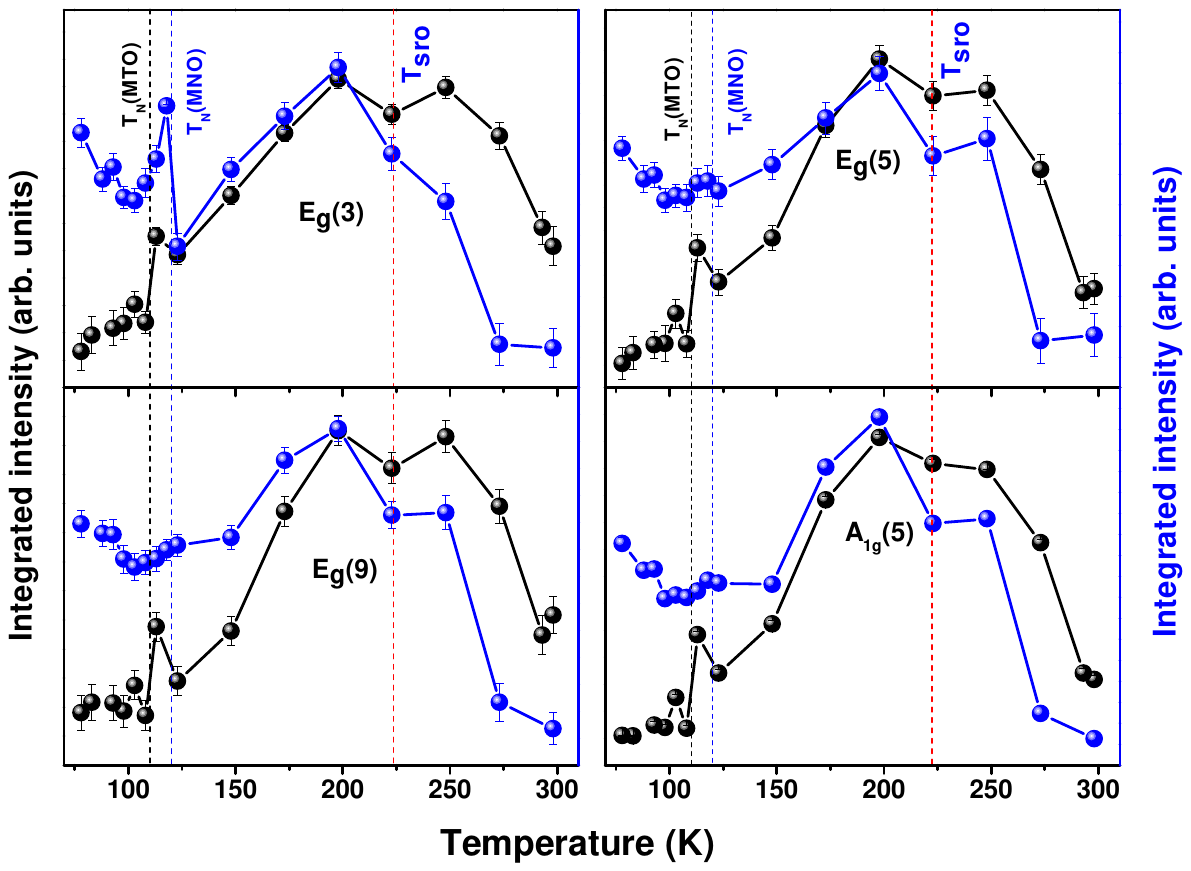}
\caption{\label{fig12}(Colour online) Temperature evolution of integrated intensity of Raman mode of MNO and MTO down to 77 K. The red dashed line indicates the magnetic short-range ordering temperature ($T_{sro}$), while blue and black dashed lines highlight the Neel temperature for MNO and MTO, respectively.} 
\end{figure*}

In an earlier low-temperature Raman study on single-crystalline MNO, no anomalous behavior in the phonon frequencies was reported, and linewidths or integrated intensities were not investigated \cite{Chen}. In contrast, the present work reveals substantial renormalization in all three Raman observables, mode frequency, linewidth, and integrated intensity. Several factors can account for the differing conclusions of the two studies. First, the previous work examined only the behavior around $T_N$, presenting relative frequency shifts in a single plot over a narrow temperature window (83-283 K). Here, we probe a comparatively broader temperature range (77-298 K) and perform a detailed, mode-resolved analysis. Since most phonons soften below $T_{sro}$ and then reverse trend and harden below $T_N,$ examining all modes together, as done previously can obscure mode-specific anomalies and diminish irregularities in their temperature evolution. Second, the earlier study employed a single-crystalline sample, which is highly susceptible to polarization-dependent intensity variations. In such cases, one Raman mode can be strongly enhanced while another becomes suppressed, resulting in improper intensity ratios between modes. This, in turn, affects the quality of Lorentzian fits and can lead to inaccuracies in the extracted mode positions. This effect is evident for the  E$_g$(10) and A$_{1g}$(4) modes: in our measurements these nearly merged modes gradually develop into distinct features at low temperatures, whereas such separation is absent in the single-crystal spectra. Distinct features were reported there only at room temperature using polarization-resolved measurements. Finally, it is well established that local laser-induced heating can significantly influence phonon energies, linewidths, and intensities. We have used approximately six times lower power to ensure minimal local heating and allowed an 8-minute stabilization time before each acquisition for thermal equilibrium. We therefore believe that the combination of low-power, thermally stabilized measurements on polycrystalline samples and careful mode-by-mode analysis enables a more accurate identification of the intrinsic renormalization of phonon frequency, linewidth, and integrated intensity reported in this study.   

Among the $A_4B_2$O$_9$ systems, only FNO exhibits a clear structural transition from trigonal \textit{P-3c1} to monoclinic \textit{C2/c} symmetry near the magnetic ordering temperature, likely driven by a Jahn–Teller (JT) distortion, as high-spin Fe$^{2+}$ is a JT-active ion \cite{Jana, Panja, Maignan}. In contrast, no long-range structural transition has been reported for the other $A_4B_2$O$_9$ compounds based on diffraction studies. A closely related situation has been observed in polar magnets (Fe,Mn)$_2$Mo$_3$O$_8$, where the Fe compound undergoes a symmetry-lowering transition concomitant with magnetic ordering, while the Mn analogue exhibits only isostructural distortions, with an incipient structural transition suggested \cite{Stanislavchuk}. In JT-active systems, orbital degeneracy is lifted through lattice distortions, promoting a lower-symmetry structure that relieves frustration and stabilizes long-range magnetic order. A similar analogy can be drawn for non–JT-active compounds through a spin-driven mechanism, widely discussed in highly frustrated pyrochlores, cubic spinels, and more recently in lower-symmetry materials \cite{Yamashita, Sushkov, Tchernyshyov, Saha, Guratinder, Lang}. In this case, a spin–Jahn–Teller (spin-JT) effect associated with magnetoelastic coupling lifts spin degeneracy and drives a transition into a lower-symmetry, frustration-relieved magnetic state \cite{Guratinder, Behr, Lang, Watanabe2, Xu}. Such spin-JT–induced distortions are typically much weaker than conventional JT distortions and may therefore remain undetected in conventional diffraction experiments \cite{Behr}. Local probes such as Raman spectroscopy are particularly sensitive to these subtle symmetry changes, revealing them through the activation of new modes, linewidth broadening, and anomalies in phonon frequency evolution \cite{Watanabe, Kirschner}. Enhanced local structural distortions combined with strong magnetic exchange interactions can significantly strengthen magnetoelastic coupling, ultimately leading to a long-range structural transition \cite{Mopoung2}. In the present Mn-based systems, this symmetry lowering is evidenced by the activation of the octahedral modes $\omega$(1) in MNO and P(1) in MTO at 148 K and 110 K, respectively. Additionally, MNO exhibits pronounced anomalies in the frequency and linewidth of multiple Raman modes around its suspected structural transition temperature $T_s \approx 148$ K. In contrast, MTO shows only modest irregularities, primarily in linewidth behavior around $T^* \approx 173$ K, which may reflect changes in short-range magnetic correlations, as its possible structural transition appears to coincide with the magnetic ordering temperature. In principle, a highly frustrated magnetic system should not stabilize long-range magnetic order even at the lowest temperatures. However, symmetry-lowering lattice distortions can lift magnetic frustration, enabling long-range magnetic order at finite temperatures \cite{Ye, Bordacs}. Within this framework, stronger frustration is expected to drive larger lattice distortions, providing a plausible explanation for the more pronounced indications of symmetry lowering in Mn-based systems compared to their Co-based counterparts.

\section{Conclusion}

In summary, our Raman spectroscopic and magnetic susceptibility studies demonstrate strong spin–phonon coupling in the spin-driven ferroelectric compounds Mn$_4$Nb$_2$O$_9$ and Mn$_4$Ta$_2$O$_9$, emerging below the short-range magnetic ordering temperature $T_{sro}\sim223$ K and undergoing further renormalization at their respective long-range ordering temperatures $T_N$. The observation of pronounced anomalies in Raman mode frequencies, and linewidths,  together with the activation of octahedral modes between $T_{sro}$ and $T_N$, provides compelling evidence for magnetoelastic coupling and suggests the onset of a low-symmetry structural distortion, more clearly resolved in MNO and closely tied to magnetic ordering in MTO. The contrasting low-temperature phonon renormalization behavior between MNO and MTO highlights the crucial role of the nonmagnetic $B$-site cation in tuning spin–lattice interactions, driven by differences in spin–orbit coupling strength and orbital hybridization between Nb$^{5+}$ (4$d$) and Ta$^{5+}$ (5$d$). Combined Raman, NMR, and DRS measurements reveal enhanced local structural distortion in Mn-based systems relative to Co analogues, while differences in electronic structure such as sub-band-gap transitions in Co compounds emphasize the sensitivity of spin-phonon coupling and magnetoelectric behavior to local structural and electronic degrees of freedom. Overall, this comprehensive study advances the microscopic understanding of spin–lattice coupling in $A_4B_2$O$_9$-type spin-driven magnetoelectric materials.

\section{Acknowledgments}
This work is supported by the National Science Foundation of China (W2432007, 42150101), the National Key Research and Development Program of China (2022YFA1402301), Shanghai Key Laboratory for Novel Extreme Condition Materials, China (No. 22dz2260800), Shanghai Science and Technology Committee, China (no. 22JC1410300). The authors acknowledge Nitin Gumber and Dr. Rajesh Pai, BARC Mumbai for ambient x-ray diffraction characterization. The authors  gratefully  acknowledge Dr. A. K. Rajarajan  BARC Mumbai  for his active support and help in sample synthesis.


\begin{references}

\bibitem{Bertaut}
E. F. Bertaut, L. Corliss, F. Forrat, R. Aleonard, and R. Pauthenet, Etude de niobates et tantalates de metaux de transition bivalents, J. Phys. Chem. Solids {\bf 21}, 234 (1961).

\bibitem{Maignan}
A. Maignan and C. Martin, Fe$_4$Nb$_2$O$_9$: A magnetoelectric antiferromagnet, Phys. Rev. B {\bf 97}, 161106(R) (2018).

\bibitem{Panja}
S. N. Panja, P. Manuel, and S. Nair, Anisotropy in the magnetization and magnetoelectric response of single crystalline Mn$_4$Ta$_2$O$_9$, Phys. Rev. B {\bf 103}, 014422 (2021).

\bibitem{Mehra}
B. S. Mehra, S. Kumar, G. Dubey, A. Shyam, A. Kumar, A. K. R., K. Singh, and D. S. Rana, Myriad of terahertz magnons with all-optical magnetoelectric functionality for eﬃcient spin-wave computing in the honeycomb magnet Co$_4$Ta$_2$O$_9$, Phys.Rev. Appl. 23, 054081 (2025).


\bibitem{Fang}
Y. Fang, Y. Q. Song, W. P. Zhou, R. Zhao, R. J. Tang, H. Yang, L. Y. Lv, S. G. Yang, D. H. Wang, and Y. W. Du, Large magnetoelectric coupling in Co$_4$Nb$_2$O$_9$, Sci. Rep. {\bf 4}, 3860 (2014).

\bibitem{Zheng}
S. H. Zheng, G. Z. Zhou, X. Li, M. F. Liu, Y. S. Tang, Y. L. Xie, M. Zeng, L. Lin, Z. B. Yan, X. K. Huang, X. P. Jiang, and
J.-M. Liu, Remarkable magnetoelectric effect in single crystals of honeycomb magnet Mn$_4$Nb$_2$O$_9$, Appl. Phys. Lett.
{\bf 117}, 072903 (2020).

\bibitem{Datta}
High-resolution neutron diffraction determination of noncollinear antiferromagnetic order in the honeycomb magnetoelectric Fe4Nb2O9Phys.Rev. B  {\bf 112}, 134439 (2025)

\bibitem{Goel}
R. Goel, K. Son, M. J. Gutmann , D. G. Oh, K. Kumar , A.Ali , S. J. Mun , D. Bhosale , G. Kim, N. Lee, Y. J. Choi, S-W. Cheong, V. Kiryukhin , and S. Choi, Ferromagnetic moment in the magnetoelectric antiferromagnet Co$_4$Ta$_2$O$_9$: Evidence of the P1 magnetic space group, Phys.Rev. Res. {\bf 7}, 043196 (2025).

\bibitem{Jana}
R. Jana, D. Sheptyakov, X. Ma, J. A. Alonso, M. Pi, A. Munoz, Z. Liu, L. Zhao, N. Su, S. Jin, X. Ma, Kai Sun, D. Chen, S. Dong, Y. Chai,
S. Li, and J. Cheng, Low-temperature crystal and magnetic structures of the magnetoelectric material Fe$_4$Nb$_2$O$_9$, Phys. Rev. B {\bf 100}, 094109 (2019).


\bibitem{Narayanan}
N. Narayanan, A. Senyshyn, D. Mikhailova, T. Faske, T. Lu, Z. Liu, B. Weise, H. Ehrenberg, R. A. Mole, W. D. Hutchison, H. Fuess, G. J. McIntyre, Y. Liu, and D. Yu, Magnetic structure and spin correlations in magnetoelectric honeycomb Mn$_4$Ta$_2$O$_9$, Phys. Rev. B {\bf 98}, 134438 (2018).


\bibitem{Deng}
G. Deng, Gang Zhao, S. Zhu, Z. Feng, W. Ren, S. Cao, ,A. Studer, and G, JMcIntyre, Spin dynamics, critical scattering and magnetoelectric coupling mechanism of Mn$_4$Nb$_2$O$_9$, New J. Phys. \textbf{24},  083007 (2022).


\bibitem{Deng2}
G. Deng, Y. Cao, W. Ren, S. Cao, A. J. Studer, N. Gauthier, M. Kenzelmann, G. Davidson, K. C. Rule, J. S. Gardner, P. Imperia, C. Ulrich, and G. J. McIntyre, Spin dynamics and magnetoelectric coupling mechanism of Co$_4$Nb$_2$O$_9$, Phys. Rev. B {\bf 97}, 085154 (2018).

\bibitem{Choi}
S. Choi, D. G. Oh, M. J. Gutmann, S. Pan, G. Kim, K. Son, J. Kim, N. Lee, S. W. Cheong, Y. J. Choi, and V. Kiryukhin, Noncollinear antiferromagnetic order in the buckled honeycomb lattice of magnetoelectric Co$_4$Ta$_2$O$_9$ determined by single-crystal neutron diffraction, Phys. Rev. B {\bf 102}, 214404 (2020).


\bibitem{Ding}
L. Ding, M.Lee, E. S. Choi, J. Zhang, Y. Wu, R. Sinclair, B. C. Chakoumakos, Y. Chai, H. Zhou, and H. Cao, Large spin-driven dielectric response and magnetoelectric coupling in the buckled honeycomb Fe$_4$Nb$_2$O$_9$, Phys. Rev. Materials {\bf 4}, 084403 (2020).


\bibitem{Zhang}
J. H. Zhang, Y. S. Tang, L. Lin, L. Y. Li, G. Z. Zhou, B. Yang, L. Huang, X. Y. Li, G. Y. Li, S. H. Zheng, M. F. Liu, M. Zeng, D. Wu, Z. B. Yan, X. K. Huang, C. Chen, X. P. Jiang, and J.-M. Liu, Electric polarization reversal and nonlinear magnetoelectric coupling in the honeycomb antiferromagnet Fe$_4$Nb$_2$O$_9$ single crystal, Phys. Rev. B \textbf{107}, 024108 (2023).

\bibitem{Panja2}
S. N. Panja, L. Harnagea, J. Kumar, P. K. Mukharjee, R. Nath, A. K. Nigam, and S. Nair, Coupled magnetic and ferroelectric states in the distorted honeycomb system Fe$_4$Ta$_2$O$_9$, Phys. Rev. B {\bf 98}, 024410 (2018). 

\bibitem{Maignan2}
A. Maignan, and C. Martin, Type-II multiferroism and linear magnetoelectric coupling in the honeycomb Fe$_4$Ta$_2$O$_9$ antiferromagnet, Phys. Rev. Mater. {\bf 2}, 091401(R) (2018).


\bibitem{Jana2}
R. Jana, A. B. Garg, B. Joseph , B. Chakraborty, I. K. A, R. Rao, Pressure-induced multiple phase transitions in the magnetoelectric Co$_4$Nb$_2$O$_9$, Phys. Rev. B {\bf 108}, 115107 (2023). 


\bibitem{Jana3}
R. Jana, R. Trivedi,  B. Joseph, A. B. Garg,  I. K. A, B. Chakraborty, M. Mukadam R. Rao, Low-temperature spin-phonon coupling and high-pressure phase transitions in the honeycomb magnetoelectric Co$_4$Ta$_2$O$_9$, Phys. Rev. B {\bf 112}, 075139 (2023).

\bibitem{Park}
K. Park, J. Kim, S.Choi, S. Fan, C. Kim, D. G. Oh, N. Lee, S.-W. Cheong, V. Kiryukhin, Y. J. Choi, D. Vanderbilt, J. H. Lee, and J. L. Musfeldt, Spin–phonon interactions and magnetoelectric coupling in Co$_4$B$_2$O$_9$ (B = Nb, Ta), Appl. Phys. Lett. {\bf 122}, 182902 (2023).

\bibitem{Ueno}
M. Ueno, T. Kurumaji, S. Kitou, M. Gen, Y. Nakamura, Y. Tokunaga, and T. Arima, Contrasting magnetothermal conductivity in sibling Co-based honeycomb-lattice antiferromagnets, Phys. Rev. B {\bf 110}, L041116 (2024).

\bibitem{Tang}
J. Tang, B. J. Lawrie, M. Cheng, Y-C Wu, H. Zhao, D. Kong, R. Lu, C.-H. Yao, Z. Gai, A-P Li, M. Li, X. Ling, Raman Fingerprints of Phase Transitions and Ferroic Couplings in van der Waals Multiferroic CuCrP$_2$S$_6$, J. Phys. Chem. Lett.  \textbf{ 16}, 4336 (2025).

\bibitem{Chen}
C. Chen, Q. Wang, H. Chen, Y. Cao, and Z. Li, Attributions of rich Raman modes and their temperature dependences in Mn$_4$Nb$_2$O$_9$ single crystals, AIP Adv. {\bf 9}, 125124 (2019).

\bibitem{Son}
Son, J., Park, B.C., Kim, C.H. et al. Unconventional spin-phonon coupling via the Dzyaloshinskii–Moriya interaction. npj Quantum Mater. \textbf{ 4}, 17 (2019).

\bibitem{Roy}
A. Prasad Roy, M. K. Chattopadhyay, Ranjan Mittal, Srungarpu N. Achary, Avesh K. Tyagi3, Manh Duc Le 6, and Dipanshu Bansal, 
Orbital ﬂuctuations and spin-orbital-lattice coupling in Bi$_2$Fe$_4$O$_9$, Phys. Rev. B \textbf{ 111}, 014310 (2025).

\bibitem{Cottam}
M. G. Cottam, D. J. Lockwood, Spin-phonon interaction in transition-metal difluoride antiferromagnets: Theory and experiment, Low Temp. Phys. 45, 78 (2019). 

\bibitem{Rao}
R. Rao, R. Selhorst, R. Siebenaller, A. N. Giordano, B. S. Conner, E. Rowe, M. A. Susner, Mode-Selective Spin–Phonon Coupling in van der Waals Antiferromagnets, Adv. Physics Res. \textbf{ 3}, 2300153 (2024).



\bibitem{Cao}
Y. Cao, M. Xiang, Z. Feng, B. Kang, J. Zhang, N. Guiblin, W. Ren, B. Dkhil, and S. Cao, Single crystal growth of Mn4Nb2O9 and its
structure-magnetic coupling, RSC Adv., \textbf{7}, 13846 (2017).

\bibitem{Solovyev} I. V. Solovyev, and T. V. Kolodiazhnyi, Origin of magnetoelectric effect in Co$_4$Nb$_2$O$_9$ and Co$_4$Ta$_2$O$_9$: The lessons learned from the comparison of first-principles-based theoretical models and experimental data, Phys Rev. B {\bf 94}, 094427 (2016).


\bibitem{Rodríguez-Hernández}
J. S. Rodríguez-Hernández , M. A. P. Gómez, O. P. Furtado, D. L. M. Vasconcelos, A. P. Ayala, and C. W. A. Paschoal, L. O. Kutelak, G. A. Lombardi, and R. D. dos Reis, CsCuCl$_3$ perovskite compound under extreme conditions, Phys. Rev. B \textbf{ 109}, 054116, 2024


\bibitem{Carlisle}
E. P. Carlisle, G. Yumnam, S. Calder, B. Haberl, Jia-Xin Xiong, M. A. McGuire, A. Zunger, R. P. Hermann, and B. A. Frandsen, Tuning the magnetic properties of the spin-split antiferromagnet MnTe through pressure Phys. Rev. B \textbf{ 112}, 014450, (2025).

\bibitem{Pawbake}
A. Pawbake, T. Pelini, I. Mohelsky, D. Jana, I. Breslavetz, C-W. Cho, M. Orlita, M. Potemski, M-A Measson, N. P. Wilson, K. Mosina, A. Soll, Z. Sofer, B. A. Piot, M. E. Zhitomirsky, C. Faugeras, Magneto-Optical Sensing of the Pressure Driven Magnetic Ground States in Bulk CrSBr, Nano Lett. \textbf{ 23}, 9587 (2023).

\bibitem{Shannon}
R. D. Shannon, Revised Effective Ionic Radii and Systematic Studies of Interatomic Distances in Halides and Chaleogenides, Acta Cryst.  \textbf{ A32}, 751 (1976).

\bibitem{Makula}
 P. Makuła,  Michał Pacia,  Wojciech Macyk, How To Correctly Determine the Band Gap Energy of Modified Semiconductor Photocatalysts Based on UV–Vis Spectra, J. Phys. Chem. Lett.  \textbf{ 9} 6814 (2018).

\bibitem{Mohanty}
P. Mohanty, S. Keshri, Estimation of band gap energy and detection of electronic transitions in UV–visible spectrum of Co$_4$Nb$_2$O$_9$, Materials Letters, \textbf{ 390}, 138409 (2025).

\bibitem{Lapina}
O. B. Lapina, D. F. Khabibulin, A. A. Shubin, and V. V. Terskikh, Practical aspects of $^{51}$V and $^{93}$Nb solid-state NMR spectroscopy and applications to oxide materials, Prog. Nucl. Magn. Reson. Spectrosc. \textbf{ 53}  128 (2008).

\bibitem{Man}
P. P. Man, Quadrupole Couplings in Nuclear Magnetic Resonance, General, Encyclopedia of Analytical Chemistry R.A. Meyers (Ed.) pp. 12224–12265 (2000).

\bibitem{Bastow}
T.J. Bastow, $^{139}$La Nuclear magnetic resonance characterisation of La$_2$O$_3$ and La$_{1-x}$Sr$_x$MO$_3$ where M = Cr, Mn or Co, Solid State Nucl. Magn. Reson.,  \textbf{ 3}  17 (1994).


\bibitem{Baek}
S-H. Baek, J. H. Lee, Y. S. Oh, K.-Y. Choi, and B. Büchner, Persistence of Ising-like easy-axis spin correlations in the paramagnetic state of the spin-1 chain compound NiTe$_2$O$_5$, Phys. Rev. B \textbf{ 104}, 214431 (2021).


\bibitem{Yu}
Y. Yu, G. Deng, Y. Cao, G. J. McIntyre, R. Li, N. Yuan, Z. Feng, Jun-Yi Ge, J. Zhang, S. Cao, Tuning the magnetic anisotropy via Mn substitution in single crystal Co$_4$Nb$_2$O$_9$, Ceram. Int. \textbf{ 45},  1093–1097 (2019).


\bibitem{Balkanski}
M. Balkanski, R. F. Wallis, and E. Haro, Anharmonic effects in light scattering due to optical phonons in silicon, Phys. Rev. B
{\bf 28}, 1928 (1983).


\bibitem{Lockwood}
D. J. Lockwood andM. G. Cottam, The spin-phonon interaction in FeF$_2$ and MnF$_2$ studied by Raman spectroscopy, J. Appl. Phys. {\bf 64}, 5876 (1988).

\bibitem{Bhadram}
V. S. Bhadram, B. Rajeswaran, A. Sundaresan, and C. Narayana, Spin-phonon coupling in multiferroic RCrO$_3$ (R-Y, Lu, Gd, Eu, Sm): A Raman study, Europhys. Lett. {\bf 101}, 17008 (2013).

\bibitem{Sohn}
 C. H. Sohn, C. H. Kim, L. J. Sandilands, N. T. M. Hien, S. Y. Kim, H. J. Park, K. W. Kim, S. J. Moon, J. Yamaura, Z. Hiroi, T. W. Noh, Strong Spin-Phonon Coupling Mediated by Single Ion Anisotropy in the All-In–All-Out Pyrochlore Magnet Cd$_2$Os$_2$O$_7$, Phys. Rev. Lett. {\bf 118}, 117201 (2017).

\bibitem{Kunwar}
H. S. Kunwar, K. Chakraborty, A. Agrawal, A. K. Yogi, B. K. De, P. Sharma, S. Mishra, S. Karmakar, D. T. Adroja, M. K. Gupta, R. Mittal, R. Venkatesh, and V. G. Sathe, Anomalous spin-lattice coupling in the quasi-one-dimensional spin-1 corrugated skew-chain antiferromagnet: Ni$_2$V$_2$O$_7$, Phys. Rev. B \textbf{ 111}, 144426 (2025).

\bibitem{Granado}
E. Granado, A. Garcia, J. A. Sanjurjo, C. Rettori, I. Torriani, F. Prado, R. Sanchez, A. Caneiro, and S. B. Oseroff, Magnetic ordering effects in the Raman spectra of La$_{1-x}$Mn$_{1-x}$O$_3$, Phys. Rev. B {\bf 60}, 11879 (1999).

\bibitem{Lee}
J. S. Lee,  T. W. Noh, J. S. Bae,  I-S. Yang, T. Takeda, and  R. Kanno, Strong spin-phonon coupling in the geometrically frustrated pyrochlore Y$_2$Ru$_2$O$_7$, Phys. Rev. B \textbf{ 69}, 214428 (2004).


\bibitem{Pal}
A. Pal, C. W. Wang, T. W. Yen, G. R. Blake, S. S. Ali, S. K. Panda, S. Kamba, Y. H. Chang, H. S. Kunwar, Y. C. Lai, Y. C. Chuang, V. Sathe, G. Senthil Murugan, R. Sankar, S. Giri, and H. D. Yang, Field-driven linear magnetoelectric coupling and entangled spin-phonon behavior in the antiferromagnetic spin-chain compound MnSb$_2$O$_4$, Phys. Rev. B \textbf{ 111}, 174453 (2025).

\bibitem{Aytan}
E. Aytan, B. Debnath, F. Kargar, Y. Barlas, M. M. Lacerda, J. X. Li, R. K. Lake, J. Shi, A. A. Balandin, Spin-phonon coupling in antiferromagnetic nickel oxide, Appl. Phys. Lett. \textbf{ 111}, 252402 (2017).

\bibitem{Poojitha}
B. Poojitha, A. Rathore, A. Kumar, and S. Saha, Signatures of magnetostriction and spin-phonon coupling in magnetoelectric hexagonal 15R-BaMnO$_3$, Phys. Rev. B {\bf 102}, 134436 (2020).

\bibitem{Laverdière}
J. Laverdière and S. Jandl A. A. Mukhin and V. Yu. Ivanov V. G. Ivanov M. N. Iliev, Spin-phonon coupling in orthorhombic RMnO$_3$ (R=Pr,Nd,Sm,Eu,Gd,Tb,Dy,Ho,Y): A Raman study, Phys. Rev. B 73, 214301 (2006).

\bibitem{Das}
S. Das, R. K. Dokala, B. Weise, P. K. Mishra, R. Medwal, R. S. Rawat, and S. Thota, Phonon scattering and magnetic manifold switching in (GdSm)CrO$_3$ Phys. Rev. Mater. 7, 084410 (2023). 

\bibitem{Rosalin}
M. Rosalin, P. Telang, S. Singh, D. V. S. Muthu, and A. K. Sood, Raman signatures of quadratic band touching state and strong spin-phonon coupling in pyrochlore iridates (Sm$_{1-x}$Bi$_x$)$_2$Ir$_2$O$_7$, Phys. Rev. B \textbf{ 108}, 035133 (2023).

\bibitem{Brown}
S. Brown and H. C. Gupta, J. A. Alonso and M. J. Martõ Ânez-Lope, Lattice dynamical study of optical modes in Tl$_2$Mn$_2$O$_7$ and In$_2$Mn$_2$O$_7$ pyrochlores, Phys. Rev. B \textbf{ 69}, 054434 (2004).


\bibitem{Lal}
S. Lal, S. J. Sebastian, S. S. Islam, M. P. Saravanan, M. Uhlarz, Y. Skourski, and R. Nath, Double magnetic transitions and exotic ﬁeld-induced phase in the triangular lattice antiferromagnets Sr$_3$Co(Nb,Ta)$_2$O$_9$, Phys. Rev. B \textbf{ 108}, 014429 (2023).

\bibitem{Watanabe}
M. Watanabe, N. Kurita, H. Tanaka, W. Ueno, K. Matsui, T. Goto, and M. Hagihala, Contrasting magnetic structures in SrLaCuSbO$_6$ and SrLaCuNbO$_6$: Spin- 1/2 quasi-square-lattice $J_1-J_2$ Heisenberg antiferromagnets, Phys. Rev. B \textbf{ 105}, 054414 (2022).

\bibitem{Mopoung}
K. Mopoung, W. Ning, M. Zhang, F. Ji, K. Mukhuti, H. Engelkamp, P. C. M. Christianen, U. Singh, J. Klarbring, S. I. Simak, I. A. Abrikosov, F. Gao, I. A. Buyanova, W. M. Chen, and Y. Puttisong, Understanding Antiferromagnetic Coupling in Lead-Free Halide
Double Perovskite Semiconductors, J. Phys. Chem. C,  \textbf{ 128}, 5313 (2024).

\bibitem{Mopoung2}
K. Mopoung, Q. Tao, F. Orlandi, K. Mukhuti, K. S. Ramsamoedj, U. Singh, S. Khamkaeo, M. Zhang, M. W. de Dreu, E. Dilmieva, E. L. Q. N. Ammerlaan, T.Ottenbros, S. Wiedmann, A. T. B., P. C. M. Christianen, S. I. Simak, J. Rosen, F. Gao, I. A. Buyanova, W. M. Chen, and Y. Puttisong, Magnetostructural Transition in Spin Frustrated Halide Double Perovskites, Chem. Mater., \textbf{ 37}, 18, 6974 (2025).

\bibitem{Sun}
Z. Sun, X. Li, H. Jin, Z. Zhao, Y. Wei , and J. Wang, Magnetic inﬂuence on phonon frequencies in two-dimensional FeOCl: Insights from spin-phonon coupling,  Phys. Rev. B \textbf{ 110}, 165415 (2024).

\bibitem{Toft}
R. Toft-Petersen, N. H. Andersen, H. Li, J. Li, W.Tian, S. L. Budko, T. B. S. Jensen, C. Niedermayer, Mark Laver, O. Zaharko, J. W. Lynn, and D. Vaknin, Magnetic phase diagram of magnetoelectric LiMnPO$_4$, Phys. Rev. B \textbf{ 85}, 224415  (2012).

\bibitem{Lado}
J. L. Lado and J. Fernández-Rossier, On the origin of magnetic anisotropy in two dimensional CrI$_3$,  2D Mater. \textbf{ 4}, 035002 (2017)

\bibitem{Xiang}
H. Xiang, C. Lee, H.-J. Koo, X. Gonga,  M.-H. Whangbo, Magnetic properties and energy-mapping analysis,  Dalton Trans. \textbf{ 42}, 823 (2013).

\bibitem{Sannigrahi}
J. Sannigrahi. Md S. Khan, M. Numan, M. Das, A. Banerjee, M. D. Le, G. Cibin, D. Adroja, and S. Majumdar, Role of crystal and magnetic structures in the magnetoelectric coupling in $CaMn_7 O_{12}$, Phys. Rev. B \textbf{ 109}, 054417 (2024).

\bibitem{Stanislavchuk}
T. N. Stanislavchuk, G. L. Pascut, A. P. Litvinchuk, Z. Liu, Sungkyun Choi, M. J. Gutmann, B. Gao, K. Haule, V. Kiryukhin, S.-W. Cheong, and A. A. Sirenko, Spectroscopic and first principle DFT+eDMFT study of complex structural, electronic, and vibrational properties of M$_2$Mo$_3$O$_8$ (M = Fe, Mn) polar magnets, Phys. Rev. B \textbf{ 102}, 115139 (2020).

\bibitem{Yamashita}
Yasufumi Yamashita and Kazuo Ueda, Spin-Driven Jahn-Teller Distortion in a Pyrochlore System, Phys. Rev. Lett. \textbf{ 85}, 4960 (2000).

\bibitem{Sushkov}
A. B. Sushkov, O. Tchernyshyov, W. Ratcliff II, S. W. Cheong, and H. D. Drew, Probing Spin Correlations with Phonons in the Strongly Frustrated Magnet $ZnCr_2O_4$, Phys. Rev. Lett. \textbf{ 94}, 137202 (2005).

\bibitem{Tchernyshyov}
O. Tchernyshyov, R. Moessner, and S. L. Sondhi, Order by Distortion and String Modes in Pyrochlore Antiferromagnets, Phys. Rev. Lett. \textbf{ 88}, 067203 (2002).

\bibitem{Saha}
Rana Saha, Francois Fauth, Maxim Avdeev, Paula Kayser, Brendan J. Kennedy, and A. Sundaresan, Magnetodielectric effects in A-site cation-ordered chromate spinels $LiMCr_4 O_8$ (M=Ga and In), Phys. Rev. B \textbf{ 94}, 064420 (2016).

\bibitem{Guratinder}
K. Guratinder, R. D. Johnson, D. Prabhakaran, R. A. Taylor, F. Lang, S.J. Blundell, L.S. Taran, Magnetoelastic Dynamics of the Spin Jahn-Teller Transition in $CoTi_2O_5$, Phys. Rev. Lett. \textbf{ 134}, 256702 (2025).

\bibitem{Lang}
Franz Lang, Lydia Jowitt , Dharmalingam Prabhakaran, Roger D. Johnson, and Stephen J. Blundell, $FeTi_2O_5$: A spin Jahn-Teller transition enhanced by cation substitution, Phys. Rev. B \textbf{ 100}, 094401  (2019).

\bibitem{Behr}
D. Behr, L. S. Taran, D. G. Porter, A. Bombardi, D. Prabhakaran, S. V. Streltsov, and R. D. Johnson, Strain-induced antiferromagnetic domain switching via the spin Jahn-Teller effect, Phys. Rev. B \textbf{ 110}, L060408, (2024).

\bibitem{Watanabe2}
T. Watanabe, S. Ishikawa, H. Suzuki, Y. Kousaka, and K. Tomiyasu, Observation of elastic anomalies driven by coexisting dynamical spin Jahn-Teller effect and dynamical molecular-spin state in the paramagnetic phase of frustrated $MgCr_2O_4$, Phys. Rev. B \textbf{ 86}, 144413 (2012).

\bibitem{Xu}
Hao-Hang Xu, Jian Liu, L. L. Tao, Xian-Jie Wang,, S. V. Streltsov, and Yu Sui, Possible spin Jahn-Teller material: Ordered pseudobrookite $FeTi_2O_5$, Phys. Rev. B \textbf{ 109}, 184430 (2024).

\bibitem{Kirschner}
F. K. K. Kirschner, R. D. J. Franz Lang, D. D. Khalyavin, P. Manue, T. Lancaster, D. Prabhakaran, and S. J. Blundell, Spin Jahn-Teller antiferromagnetism in $CoTi_2O_5$, Phys. Rev. B \textbf{ 99}, 064403 (2019).

\bibitem{Ye}
F. Ye, Y. Ren, Q. Huang, J. A. Fernandez-Baca, Pengcheng Dai, J. W. Lynn, and T. Kimura, Spontaneous spin-lattice coupling in the geometrically frustrated triangular lattice antiferromagnet $CuFeO_2$, Phys. Rev. B \textbf{ 73}, 220404(R) (2006).

\bibitem{Bordacs}
S. Bordacs, D. Varjas, I. Kezsmarki, G. Mihaly, L. Baldassarre, A. Abouelsayed, C.A. Kuntscher, K. Ohgushi, and Y. Tokura, Magnetic-Order-Induced Crystal Symmetry Lowering in $ACr_2O4$ Ferrimagnetic Spinels, Phys. Rev. Lett. 103, 077205 (2009).

\end{references}
\end{document}